%% file: ms.tex
\documentclass[12pt,preprint]{aastex}
\shorttitle{SN~2005bf}
\shortauthors{CSP et al.}

\received{2005 September 16}
\begin{document}

\title{SN~2005bf: A {\bf Possible} Transition Event Between Type Ib/c
  Supernovae and Gamma-Ray Bursts}

\author{Gast\'on Folatelli\altaffilmark{1},
Carlos Contreras\altaffilmark{1},
M.~M.~Phillips\altaffilmark{1},
S.~E.~Woosley\altaffilmark{2},
Sergei Blinnikov\altaffilmark{3},
Nidia Morrell\altaffilmark{1,12}, 
Nicholas~B.~Suntzeff\altaffilmark{4},
Brian~L.~Lee\altaffilmark{5},
Mario Hamuy\altaffilmark{1,10},
Sergio Gonz\'alez\altaffilmark{1},
Wojtek Krzeminski\altaffilmark{1},
Miguel Roth\altaffilmark{1},
Weidong Li\altaffilmark{6},
Alexei~V.~Filippenko\altaffilmark{6},
Ryan~J.~Foley\altaffilmark{6},
W.~L.~Freedman\altaffilmark{7},
Barry~F.~Madore\altaffilmark{7,8},
S.~E.~Persson\altaffilmark{7},
David Murphy\altaffilmark{7},
Samuel Boissier\altaffilmark{7},
Gaspar Galaz\altaffilmark{9},
Luis Gonz\'alez\altaffilmark{10},
P.~J.~McCarthy\altaffilmark{7},
Andrew McWilliam\altaffilmark{7}, and
Wojtek Pych\altaffilmark{11}}

\altaffiltext{1}{Las Campanas Observatory, Carnegie Observatories,
  Casilla 601, La Serena, Chile.}
\altaffiltext{2}{Department of Astronomy, University of California,
  Santa Cruz, CA 95064.} 
\altaffiltext{3}{MPA D-85741 Garching, Germany, and ITEP 117218 Moscow, Russia.}
\altaffiltext{4}{Cerro Tololo Inter-American Observatory, Casilla 603,
  La Serena, Chile.} 
\altaffiltext{5}{Department of Astronomy and Astrophysics, University
  of Toronto, 60 St. George St., Toronto, Ontario, Canada M5S 3H8.}
\altaffiltext{6}{Department of Astronomy, University of California,
  Berkeley, CA 94720-3411.}
\altaffiltext{7}{Observatories of the Carnegie Institution of
  Washington, 813 Santa Barbara St., Pasadena, CA 91101.}
\altaffiltext{8}{Infrared Processing and Analysis Center, Caltech/Jet
  Propulsion Laboratory, Pasadena, CA 91125.}
\altaffiltext{9}{Departamento de Astronom\'{\i}a y Astrof\'{\i}sica,
  Pontificia Universidad Cat\'olica de Chile, Casilla 306, Santiago
  22, Chile.} 
\altaffiltext{10}{Universidad de Chile, Departamento de Astronom\'{\i}a,
  Casilla 36-D, Santiago, Chile.}
\altaffiltext{11}{Copernicus Astronomical Center, Bartycka 18,
  PL-00716 Warszawa, Poland.} 
\altaffiltext{12}{Member of the Carrera del Investigador
  Cient\'{\i}fico, CONICET, Argenina; on leave from Facultad de
  Ciencias Astron\'omicas y Geof\'{\i}sicas, Universidad Nacional de
  La Plata.} 

%\email{gfolatelli, ccontreras, mmp, nmorrell, miguel, 
%sgonzalez, wojtek@lco.cl \\
%woosley@ucolick.org \\
%seb@MPA-Garching.MPG.DE \\
%nsuntzeff@noao.edu \\
%mhamuy, lgonzalez@das.uchile.cl \\
%blee@astro.utoronto.ca \\
%weidong@astron.berkeley.edu \\
%alex@astron.berkeley.edu \\
%wendy, persson, david, pmc2, andy, boissier@ociw.edu \\
%barry@ipac.caltech.edu \\
%ggalaz, elabbe@astro.puc.cl \\
%weinberger, akir@dtm.ciw.edu \\
%pych@camk.edu.pl}

\begin{abstract}
  \noindent We present $u'g'r'i'BV$ photometry and optical
  spectroscopy of the Type~Ib/Ic SN~2005bf covering the first
  $\sim$100 days following discovery.  The $u'g'BV$ light curves displayed
  an unprecedented morphology among Type~Ib/Ic supernovae, with an
  initial maximum some 2 weeks 
  after discovery, and a second, main maximum about 25 days after that.
  The bolometric light curve indicates 
  that SN~2005bf was a remarkably luminous event, radiating at least 
  $6.3\times 10^{42}$ erg s$^{-1}$ at maximum light, and a total of 
  $2.1\times 10^{49}$ erg during the first 75 days after the explosion. 
  Spectroscopically, SN~2005bf underwent a unique transformation 
  from a Type~Ic-like event at early times to a typical Type~Ib supernova 
  at later phases. The initial maximum in $u'g'BV$ was accompanied by the 
  presence in the spectrum of high velocity ($> 14,000$ km s$^{-1}$) 
  absorption lines of \ion{Fe}{2}, \ion{Ca}{2}, and \ion{H}{1}. The 
  photospheric velocity derived from spectra at early epochs was below 
  10,000 km s$^{-1}$, which is unusually low compared with ordinary Type~Ib 
  supernovae.
  We describe one-dimensional computer simulations which attempt to account
  for these remarkable properties. The most favored model is that of a very 
  energetic ($2 \times 10^{51}$ erg), asymmetric explosion of a massive 
  (8.3 M$_\odot$) Wolf-Rayet WN star that had lost most of 
  its hydrogen envelope.  We speculate that an unobserved relativistic jet 
  was launched producing a two-component 
  explosion consisting of (1) a polar explosion containing a small fraction 
  of the total mass and moving at high velocity, and (2) the explosion of the 
  rest of the star.  At first, only the polar explosion is observed, 
  producing the initial maximum and the high velocity absorption-line 
  spectrum resembling a Type~Ic event.  At late times, this fast-moving 
  component becomes optically thin, revealing the more slowly moving explosion 
  of the rest of the star and transforming the observed spectrum to that of 
  a typical Type~Ib supernova.  If this scenario is correct, then SN~2005bf 
  is the best example to date of a transition object between normal 
  Type~Ib/Ic supernovae and gamma-ray bursts.
\end{abstract}


\section{INTRODUCTION}
\label{sec:intro}

\noindent After years of relative neglect, Type~Ib/Ic supernovae 
\citep[SNe; see][for a discussion of SN spectral
  types]{filippenko97} 
are today the object of considerable observational and theoretical 
attention.  This boom is largely due to the discovery that certain
gamma-ray bursts (GRBs) are associated with highly energetic Type~Ic
events commonly referred to as ``hypernovae''
\citep[e.g., see][and references therein]{stanek05}. Note, however,
that this term has been
expanded by some authors to include essentially any Type~Ic SN with 
inferred high luminosity or unusually broad lines, whether or not it 
was associated with a GRB.

\noindent It is universally agreed that Type~Ib/Ic SNe, as well as the 
transitional Type~IIb SNe, are produced by the core collapse
of a massive star which has lost most or all of its hydrogen 
envelope before exploding \citep[e.g., see][and references
  therein]{filippenko05}.  Taken together as a class, the light 
curves of Type~Ib/Ic/IIb SNe display a significant range in luminosities 
and morphologies, although there are subclasses of events which are 
relatively homogeneous in their properties \citep{clocchiatti97}.  
Unfortunately, only a handful of Type~Ib/Ic SNe have been observed 
significantly before maximum \citep[e.g.,][]{foley03}; thus, 
little is known of their 
photometric and spectroscopic properties at the earliest epochs 
following outburst.

\noindent At the Las Campanas Observatory (LCO), 
we have embarked on a multi-year
observational program --- the Carnegie Supernova Program (CSP) --- to 
obtain follow-up photometry and spectroscopy
of SNe of all types \citep[see][hereafter Paper I]{hamuy06}. 
One of the many goals of this project is 
to obtain uniform, high-precision observations of nearby 
Type~Ib/Ic/IIb SNe in order to improve our knowledge of the optical and
near-infrared
characteristics of these explosions. Highest priority is given to
SNe discovered soon after outburst, since it is at these early
epochs where the largest range in photometric and spectroscopic
properties is likely to be observed.

\noindent One of the Type~Ib/Ic events to be studied in 
detail as part of the CSP was SN~2005bf, which was independently 
discovered \citep{monard05} on 2005 April 6 (UT dates are used
throughout this paper) by L. A. G. Monard 
at the Bronberg Observatory in South Africa, and by M. Moore 
and W. Li during the Lick Observatory Supernova Search
\citep[LOSS;][]{li00,filippenko01,filippenko05} .
The SN was located 11$''$.7 east and 32$''$.6 south
of the nucleus of MCG +00-27-5, an SB galaxy with a heliocentric
recession velocity 
of 5670 km s$^{-1}$ \citep{falco99}. According to Monard, nothing was
visible at the position of the SN on an unfiltered image taken on 
March 12 with a limiting magnitude of 18.5. Moore and Li report 
no detection to a limiting unfiltered magnitude of 19.5 on an 
image from March 15, and a marginal detection at 18.8 on March 30. It
was likely, therefore, that the explosion occurred between March
15 and 29. 

\noindent Initial CSP spectroscopy revealed that SN~2005bf was a
Type~Ic event \citep{morrell05}, and optical photometry confirmed
that the SN had, indeed, been caught on the rising part of the 
light curve.  An intensive program of follow-up photometric and 
spectroscopic observations was initiated which unexpectedly 
revealed an unusual preliminary rise and fall in the $u', g', B,$ 
and $V$ light curves preceding by $\sim$25 days the main maxima 
in these bands \citep{hamuy05}.  Spectra obtained just before the
time of the main maximum also revealed that SN~2005bf had transformed into 
a Type~Ib or transitional Type~IIb event with strong \ion{He}{1} 
lines \citep{wang05}.

\noindent In this paper, we present the CSP observations of 
SN~2005bf, and discuss these in the context of possible models of
the star that produced this peculiar and, thus far, unique explosion.

\section{PHOTOMETRY}
\label{sec:phot}

\begin{figure}
\plotone{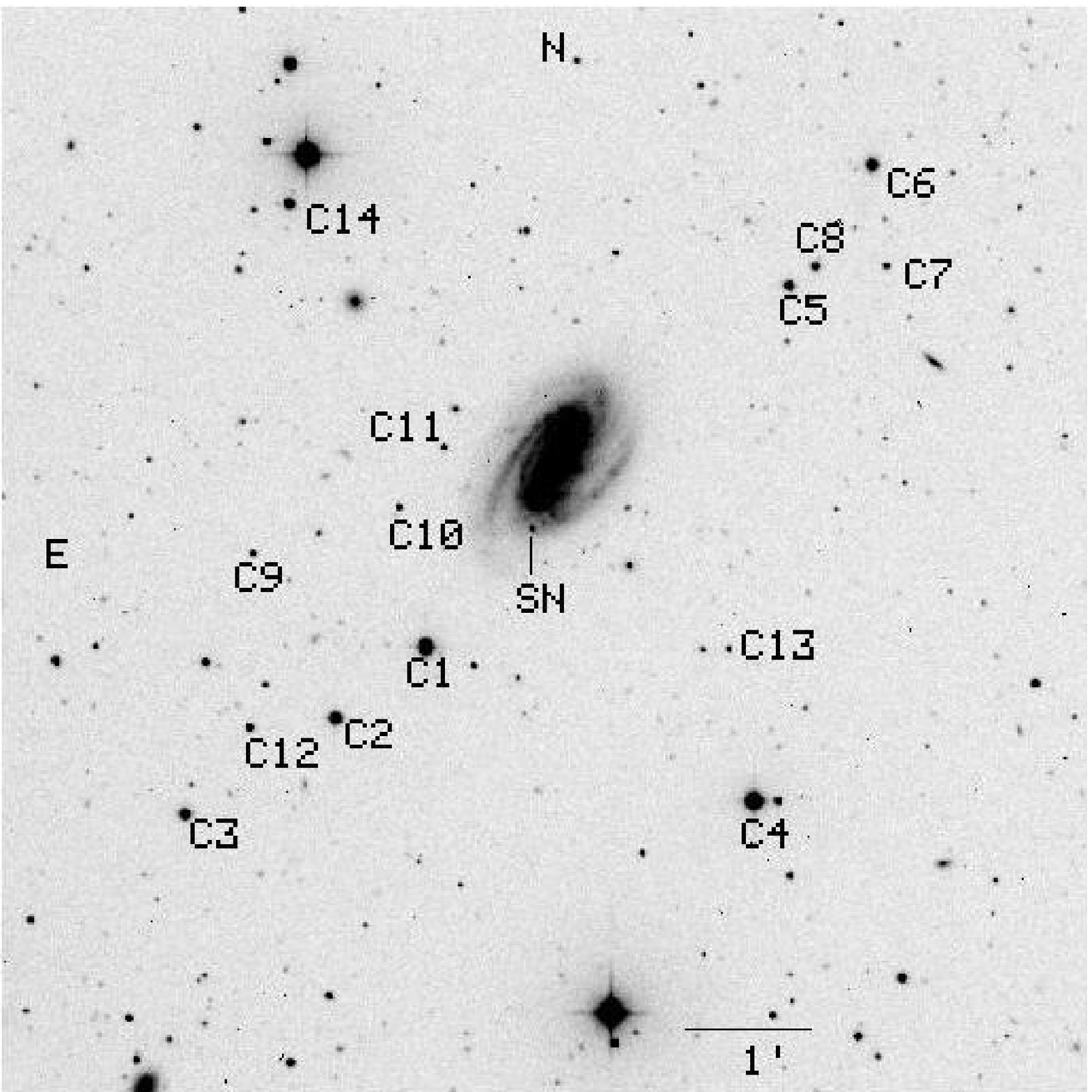}
\caption{The field of SN~2005bf observed with the Swope 1-m telescope at
  LCO and a $V$ filter. North is up and east is to the left. The
  supernova is marked to 
  the southeast of the host-galaxy nucleus. Fourteen comparison stars
  used to derive differential photometry of the SN are labeled. The
  image scale is shown near the bottom.\label{fig:fc}}   
\end{figure}

\noindent Most of our optical photometry was obtained 
with the Swope 1-m telescope at LCO, using
a SITe CCD and a set of Sloan Digital Sky Survey (SDSS) $u'g'r'i'$ and
Johnson $BV$ filters \citep{fukugita96,bessell90}. We read a section
of 1200$\times$1200 pixels 
from the CCD, which, at a scale of $0.''435$ pixel$^{-1}$, yielded a
field of view of 8.$'$7$\times$8.$'$7. Typical image quality ranged
between 1$''$ and 2$''$ (FWHM).
A photometric sequence of comparison stars in the SN field was
calibrated with the 
Swope telescope from observations of standard stars of
\citet{landolt92,smith02} during four photometric
nights. Figure~\ref{fig:fc} shows the SN field and the selected comparison
stars. Table~\ref{tab:stds} lists the average $u'g'r'i'BV$
magnitudes derived for these stars.
SN magnitudes in the standard SDSS+Johnson system were obtained
differentially relative to the comparison stars using
point-spread-function (PSF) photometry. On every image, a PSF was
fitted to the SN and comparison stars within a radius of $3''$. We
refer to Paper I for further details about the instrument and
measurements.

\noindent Five unfiltered LOSS images obtained with the 0.8-m
Katzman Automatic Imaging Telescope
\citep[KAIT;][]{filippenko01,filippenko05} were included in our
analysis because they allow us to study the very early stages of the
SN. After some experimentation 
we found that the unfiltered instrumental magnitudes of the local standards
could be satisfactorily transformed with a simple additive zero-point
to the $r'$ standard system with a dispersion $\lesssim$0.05 mag. The
comparison of the resulting LOSS magnitudes of the SN on April 15
(JD~2,453,475; $r'$ = 17.419 mag) and 16
(JD~2,453,476; $r'$ = 17.426 mag) with those
obtained with the Swope telescope on the same nights ($r'$ = 17.409, and
17.406, respectively) confirms that the effective wavelength of the
$r'$ filter is a good match to that of the KAIT unfiltered bandpass. The
first LOSS observation obtained on March 15 (JD~2,435,445),
where the SN is not detectable, was used to derive a lower limit to the
SN magnitude of $r'> 20.0$ mag. 

\noindent A 17-day gap in the $r'$ and $i'$ observations with the
Swope telescope was partially filled with an $r'$ image and a
spectrum obtained on May 7 (JD~2,453,497) using the Low
Dispersion Survey Spectrograph 3 (LDSS-3) at the Clay 6.5-m 
telescope at LCO. For this purpose, we used the filter response
functions of $r'$ and $i'$ given by \citet{smith02} to compute a synthetic
$(r'-i')$ color from the spectrum, and thence to derive a synthetic $i'$
magnitude from the observed $r'$ magnitude. 

\begin{figure}
\plotone{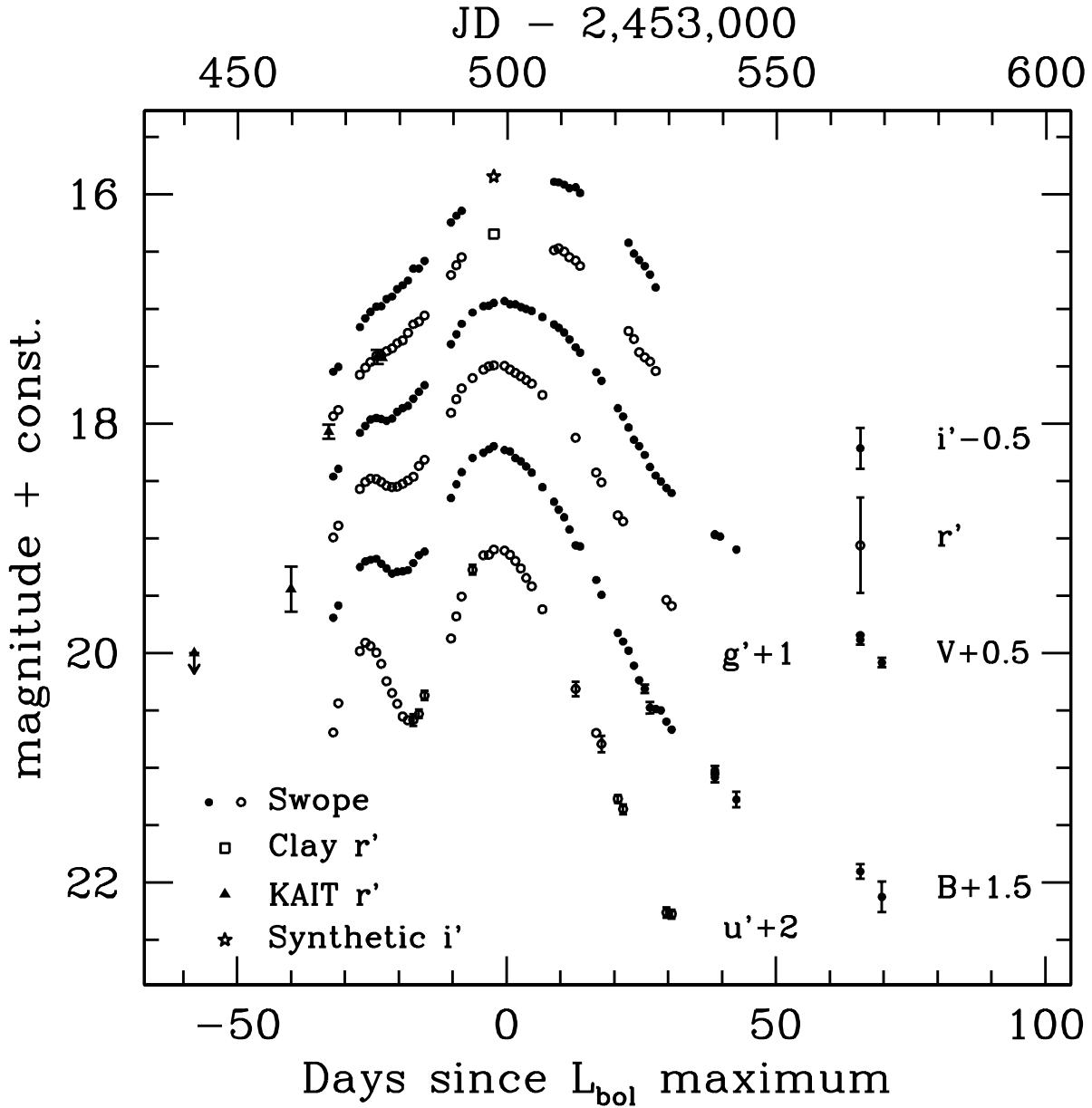}
\caption{Observed $u'g'r'i'BV$ light curves of SN~2005bf from Swope
  ({\em open and closed circles\,}), KAIT ({\em closed triangles\,}),
  and Clay 
  ({\em open square\,}). A star symbol marks a
  synthetic $i'$ magnitude obtained from the LDSS-3 spectrum of May
  7 ($\sim$JD~2,453,497). The arrow marks an $r'$ lower limit from a
  KAIT image obtained on 
  March 15 ($\sim$JD~2,453,445). Unless explicitly drawn, the error
  bars are smaller than the symbols. The time axes are given in the
  observer's frame. For clarity, the magnitudes in each band have been 
  shifted by an arbitrary constant.\label{fig:lcs}}  
\end{figure}

\noindent The resulting $u'g'r'i'BV$ magnitudes of SN~2005bf and their
uncertainties are listed in Table~\ref{tab:phot}. A minimum
uncertainty of 0.015 mag was assumed for a single measurement based on
the typical scatter in the transformation from instrumental to standard
magnitudes of bright stars (Paper I). Figure~\ref{fig:lcs} shows
the corresponding light curves. We computed K-corrections for the
$g'r'i'BV$ filters (the $u'$ passband lies beyond the coverage of our
spectra) using the available spectra. Given the small redshift of the
SN, the corrections proved small, within $\pm$0.02 mag  
in all cases. We therefore decided to neglect them in our analysis.  

\noindent Errors in our photometry could potentially arise from
a poor subtraction of underlying host-galaxy light.
Fortunately, SN~2005bf was located in an outer region of the
galaxy (see Figure~\ref{fig:fc}). Inspection of the pre-discovery template 
LOSS images reveals no strong background variations at the location
of the SN within a radius of $\sim$3$''$ or in the sky annulus.
If the contamination were large, the seeing changes from epoch to epoch would
cause a large variation in such contamination and, consequently, a large
scatter in the light curves. The small scatter seen in the light curves 
(typically 0.01--0.03 mag) is an indication that the contamination must
be very small.

\noindent As mentioned in \S~\ref{sec:intro}, the $u'g'BV$ 
light curves of SN~2005bf show a unique
behavior consisting of a first maximum around April 13 (JD~2,453,474),
a subsequent decline lasting several days, a second rise which brings
the SN to a main maximum around May 8 (JD~2,453,498), and finally
a 30-day fast decline phase followed by a slower decline 
phase. Although the $r'i'$ light curves do not show this double maximum,
they do show a shoulder around the time of the first maximum. The
other remarkable 
feature is the long time ($\sim$40 days) in all filters the SN took to
reach the main maximum. There is a large range in light-curve width
among Type~Ib/Ic SNe \citep{hamuy04}, with the ``hypernovae'' SN~1997ef
and SN~1998bw having two of the broadest ones, but even these objects
took between 15 and 20 days to reach maximum light
\citep{galama98,iwamoto00}. Table~\ref{tab:lpar} lists the 
dates of maximum and peak magnitudes for SN~2005bf in $u'g'r'i'BV$,
and in the bolometric light curve (see \S~\ref{sec:bol}).

\begin{figure}
\plotone{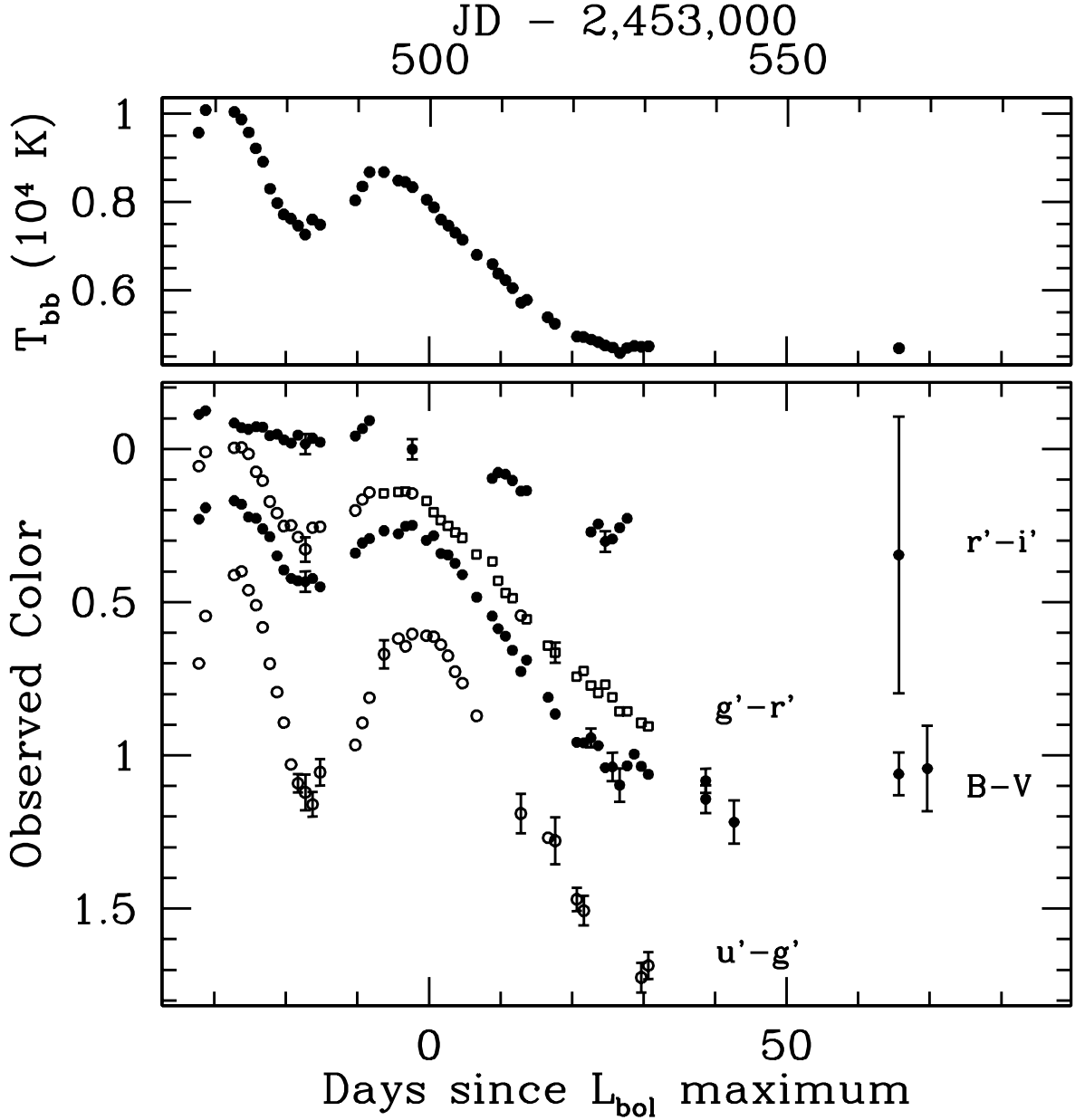}
\caption{(top) Temperature $T_{bb}$ from the blackbody fits to the
  $B$ through $i'$ monochromatic fluxes. (bottom) Observed color 
  evolution for ($u'-g'$), ($B-V$), ($g'-r'$), and ($r'-i'$). 
  Circles mark colors computed directly from measurements in both
  bands while squares are used when one of the light curves was
  interpolated in time. Unless explicitly drawn, the error
  bars are smaller than the symbols. The time axes are given in the
  observer's frame. No reddening
  correction was applied.\label{fig:tcol}} 
\end{figure}

\noindent Figure~\ref{fig:tcol} shows the SN colors along with the
temperature derived from blackbody (BB) fits to the $g'r'i'BV$
magnitudes as described in \S~\ref{sec:bol}. All of the 
color curves track the temperature evolution
very closely. We observe a temperature rise 
reaching 10,000 K at the time of the first maximum, a cooling phase for
a period of 10 days which brings the temperature down to 7300 K, a
subsequent reheating period of 13 days reaching 8700 K close to the
time of maximum luminosity, followed by a steady second cooling phase
past maximum light. 

\noindent In addition to the optical data, $YJHK_s$ photometry was 
obtained on three epochs using the Wide Field Infrared Camera
\citep[WIRC;][]{persson02} and the Persson Auxiliary Nasmyth Infrared Camera
\citep[PANIC;][]{martini04} mounted
on the duPont 2.5-m, and the Baade 6.5-m telescopes at LCO,
respectively. Two comparison stars in the SN field were calibrated
using observations of standard stars of \citet{persson98} obtained on May 21 
(JD~2,453,511) with the Baade 6.5-m telescope. Table~\ref{tab:nirst}
lists the $YJHK_s$ magnitudes for the comparison stars. SN magnitudes
were computed differentially relative to the comparison stars with an
aperture of $2''$ (see Paper I for details). The $YJHK_s$
photometry for SN 2005bf is listed in Table~\ref{tab:nir}. The quoted
uncertainties assume a minimum error of 0.02 mag on an individual
measurement (Paper I). 

\section{SPECTROSCOPY}
\label{sec:spec}

\noindent A total of nine optical spectra were obtained using the 
long-slit grism mode of the Wide Field Reimaging CCD Camera (WFCCD) at the
duPont 2.5-m telescope and the Low-Dispersion Survey Spectrograph
\citep[LDSS-3;][]{allington94} spectrograph at the Clay 6.5-m 
telescope, both at LCO. An additional spectrum was obtained with the
Low Resolution Imaging Spectrometer \citep[LRIS;][]{oke95} on the Keck-I
10-m telescope. Table~\ref{tab:spec} gives a journal of the
spectroscopic observations. 

\noindent In general, the spectra were obtained using a narrow
($\sim$1--2$''$) slit oriented near the parallactic angle
\citep{filippenko82}. All of the WFCCD spectra were obtained with a
single instrumental setup, without order-blocking filters, and are
thus subject to second-order contamination effects at wavelengths
longer than 7000\AA. The LDSS-3 and LRIS spectra were instead obtained
with two instrumental setups (blue and red channels) and using
order-blocking filters.
The spectrum from May 31 (JD~2,543,521) was
obtained through a 
wider ($8.''65$) slit. All the spectra were wavelength and flux
calibrated using arc-lamp and flux-standard observations
\citep{hamuy94}. In some cases, a smooth-spectrum star selected from
Table~4 of \citet{bessell99} was observed to correct for telluric
absorption features (see Paper I).

\noindent Additionally, a spectrum of the underlying host galaxy near
the location of the SN was extracted from the LDSS-3 observation of
May 7 (JD~2,453,497). Several emission (H$\alpha$ and
[\ion{N}{2}] $\lambda$$\lambda$6548, 6583) and absorption (\ion{Na}{1}
D1 \& D2, and \ion{Ca}{2} H \& K) lines were identified, which allowed
us to derive an average recession velocity at the location of the SN
of 5496 $\pm$ 27 km s$^{-1}$. This is somewhat lower than the
recession velocity of 5670 $\pm$ 8 km s$^{-1}$ given by
\citet{falco99} for the host galaxy, a difference  
probably caused by the rotation of the galaxy. The former velocity
was adopted to shift the observed wavelength axis of the spectra to
the SN rest frame.

\noindent Figure~\ref{fig:spec} shows the series of SN spectra sorted in
time. (The LDSS-3 spectrum on JD~2,453,502.48 is excluded, to avoid 
crowding; it spans only 6057--10,000~\AA, in any case.) 
The main spectral features are labeled with the ions that produce
them. The positions of the two main telluric absorption features are
marked with an Earth symbol. The numbers in parenthesis give the
epoch of each spectrum in SN rest-frame days since the time of maximum
bolometric luminosity (JD~2,453,499.8, see \S~\ref{sec:bol}). We
will hereafter use these epochs to label the spectra.

\section{ANALYSIS}
\label{sec:anal}

\subsection{Spectroscopic Analysis}
\label{sec:spanal}

\begin{figure}
\plotone{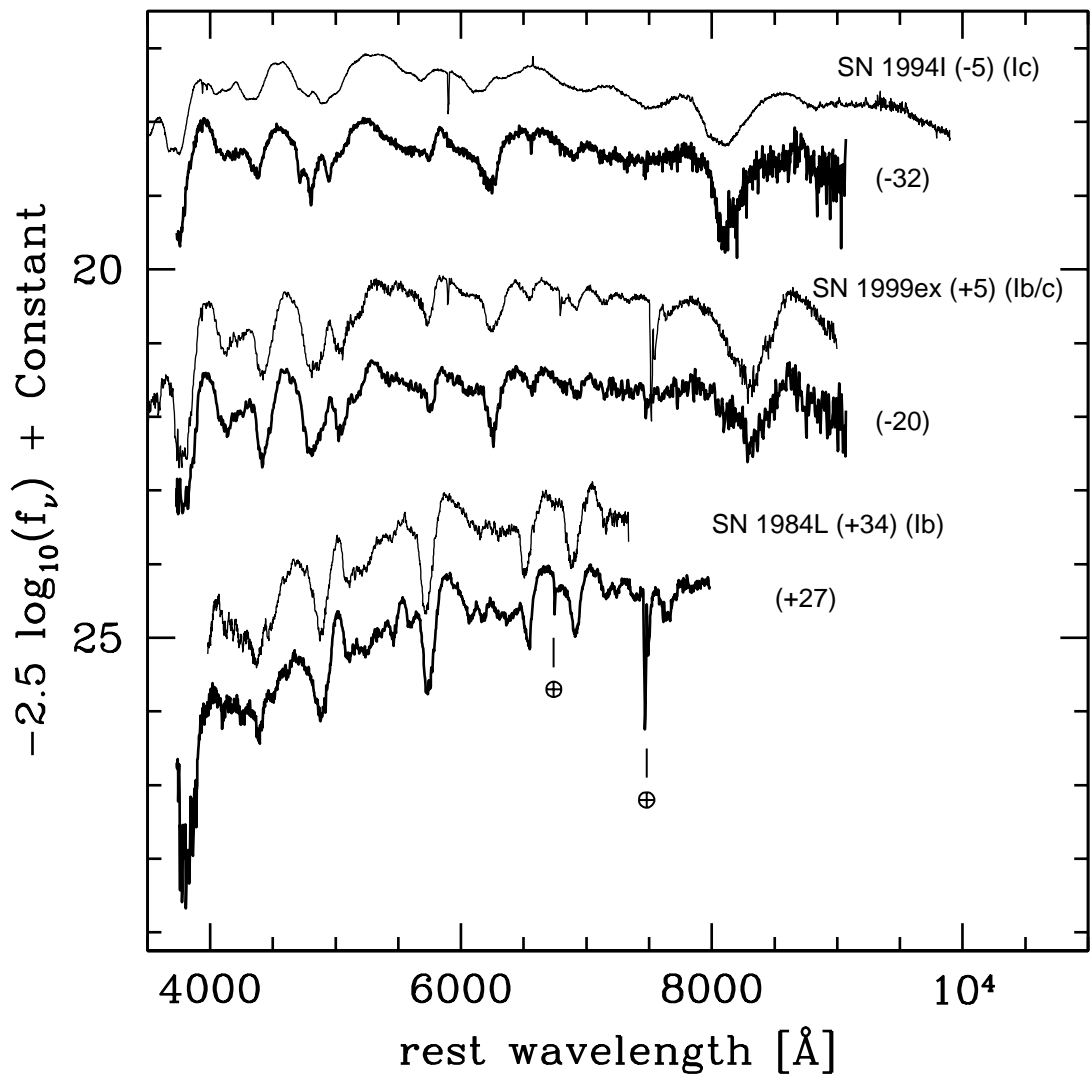}
\caption{Comparison of spectra of SN~2005bf with those of other
  Type~Ib and Type~Ic SNe. At the top, the spectrum from day $-32$
  compared with that of the 
  Type~Ic SN~1994I five days before maximum light
  \citep{filippenko95}. At the middle, the
  spectrum from day $-20$ and that of the Type~Ib/Ic SN~1999ex five days
  after maximum light \citep{hamuy02}. At the bottom, the spectrum
  from day $+27$ and that of the Type~Ib SN~1984L 34 days after maximum
  light \citep{harkness87}. Telluric features are marked with an 
  Earth symbol ($\oplus$).\label{fig:spcomp}} 
\end{figure}

\noindent SN~2005bf was classified by \citet{morrell05} as a Type~Ic
event due to the absence of strong \ion{H}{1} and \ion{He}{1} lines 
which are the distinguishing features of Type~II and Ib SNe 
\citep[e.g.,][]{filippenko97}, respectively. Although the initial
spectra of 
SN~2005bf showed a prominent absorption at $\sim$6250~\AA~that
could be attributed to Si~II $\lambda$6355, 
it was not nearly as strong as is observed in typical 
Type~Ia SNe, nor was there any evidence in the spectra for the
\ion{S}{2} $\lambda$$\lambda$5454, 5640 feature which is also prominent
around maximum light in Type~Ia SNe. The two topmost spectra shown in
Figure~\ref{fig:spcomp} compare SN~2005bf (day $-32$) and the
Type~Ic SN~1994I five days before maximum light
\citep{filippenko95}. The comparison shows that at
early times, SN~2005bf could be classified as a Type~Ic SN. A similar
comparison of SN~2005bf with
an early-time spectrum of SN~1987M led \citet{modjaz05} to 
independently classify SN~2005bf as a Type~Ic
event. As SN~2005bf evolved, however, the \ion{He}{1} lines grew in
strength, and as shown at the bottom of Figure~\ref{fig:spcomp}, the
spectra from day $+21$ and later became almost 
identical to those of the Type~Ib SN~1984L. This
transformation was also noticed by \citet{wang05}. Interestingly, as
shown in the middle of Figure~\ref{fig:spcomp}, the
spectrum from day $-20$ (and also those from days $-2$ and $+2$)
was remarkably similar to spectra of the intermediate Type~Ib/Ic
SN~1999ex obtained around maximum light \citep{hamuy02}.  

\begin{figure}
\epsscale{0.92}
\plotone{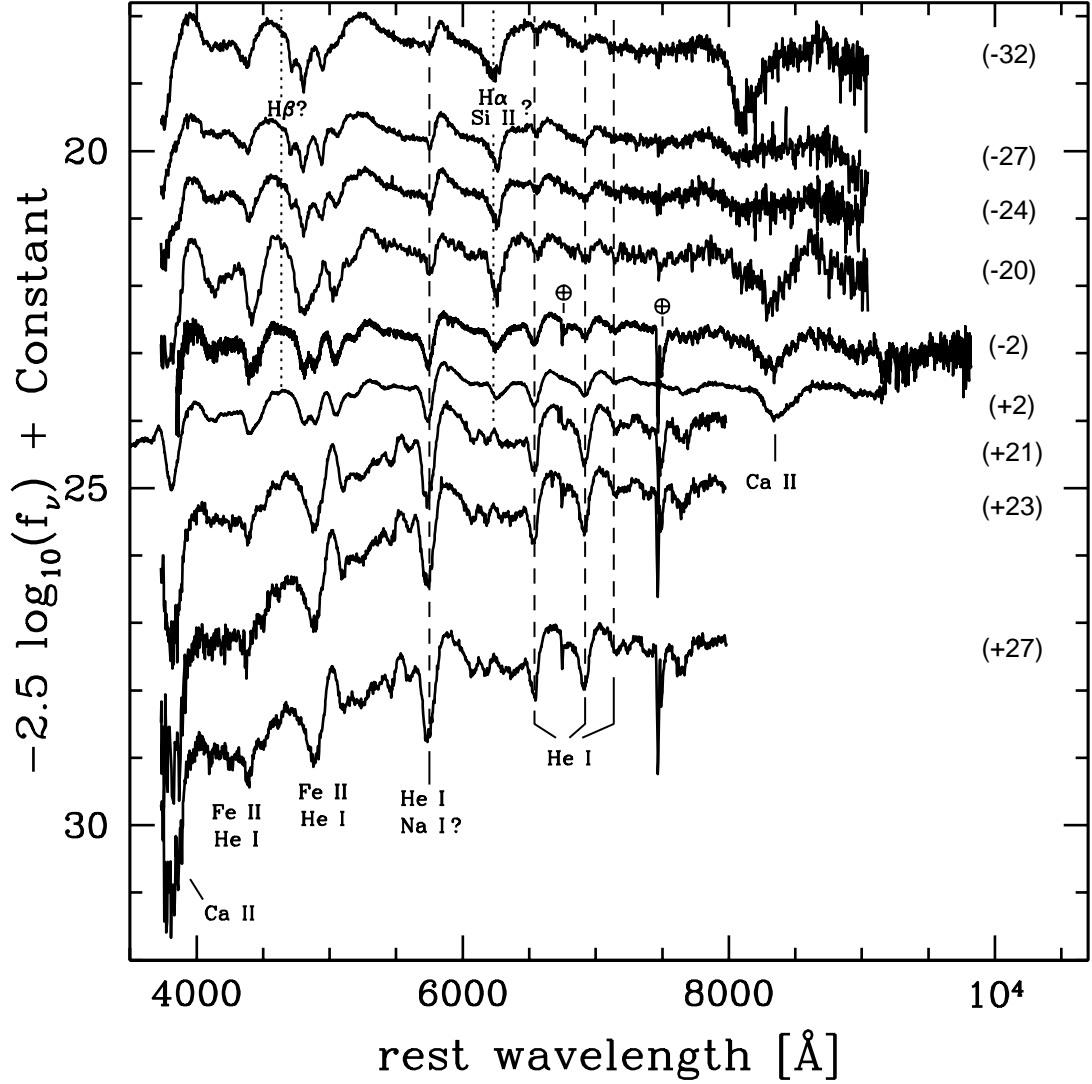}
\caption{Spectroscopic evolution of SN~2005bf. Each spectrum is
  presented on a 
  logarithmic scale, shifted by an arbitrary constant. The wavelength
  of the spectra was shifted to the SN rest frame using a redshift of
  $z = 0.01833$. The labels in parentheses to 
  the right of each spectrum indicate the epoch in rest-frame days since
  $L_{\mathrm {bol}}$ maximum (JD $=$ 2,453,499.8). The spectrum from day
  $-20$ was smoothed by averaging over five pixels. The name 
  of the ions responsible for some of the main features are given. The
  dotted vertical lines mark the wavelength at which the minimum of the
  H$\alpha$ and H$\beta$ lines would appear if hydrogen were present at
  $\sim$14,500 km s$^{-1}$. The dashed vertical lines mark the
  absorptions due to \ion{He}{1} $\lambda$$\lambda$
  5876, 6678, 7065, 7281, at an expansion velocity of $\sim$6000 km 
  s$^{-1}$. Telluric features are marked with an 
  Earth symbol ($\oplus$).\label{fig:spec}}
\end{figure}

\noindent Figure~\ref{fig:spec} indicates that the
\ion{He}{1} $\lambda$$\lambda$5876, 6678, 7065 lines were
present, albeit weakly, in the first spectra of SN~2005bf,
and grew steadily in strength during the period covered by our
observations.  The expansion velocities as measured from the
absorption minima of these features were nearly constant at a
value of $\sim$8000--10,000 km s$^{-1}$ until maximum light and
dropped to $\sim$6000  km s$^{-1}$ after that.

\begin{figure}
\epsscale{.70}
\plotone{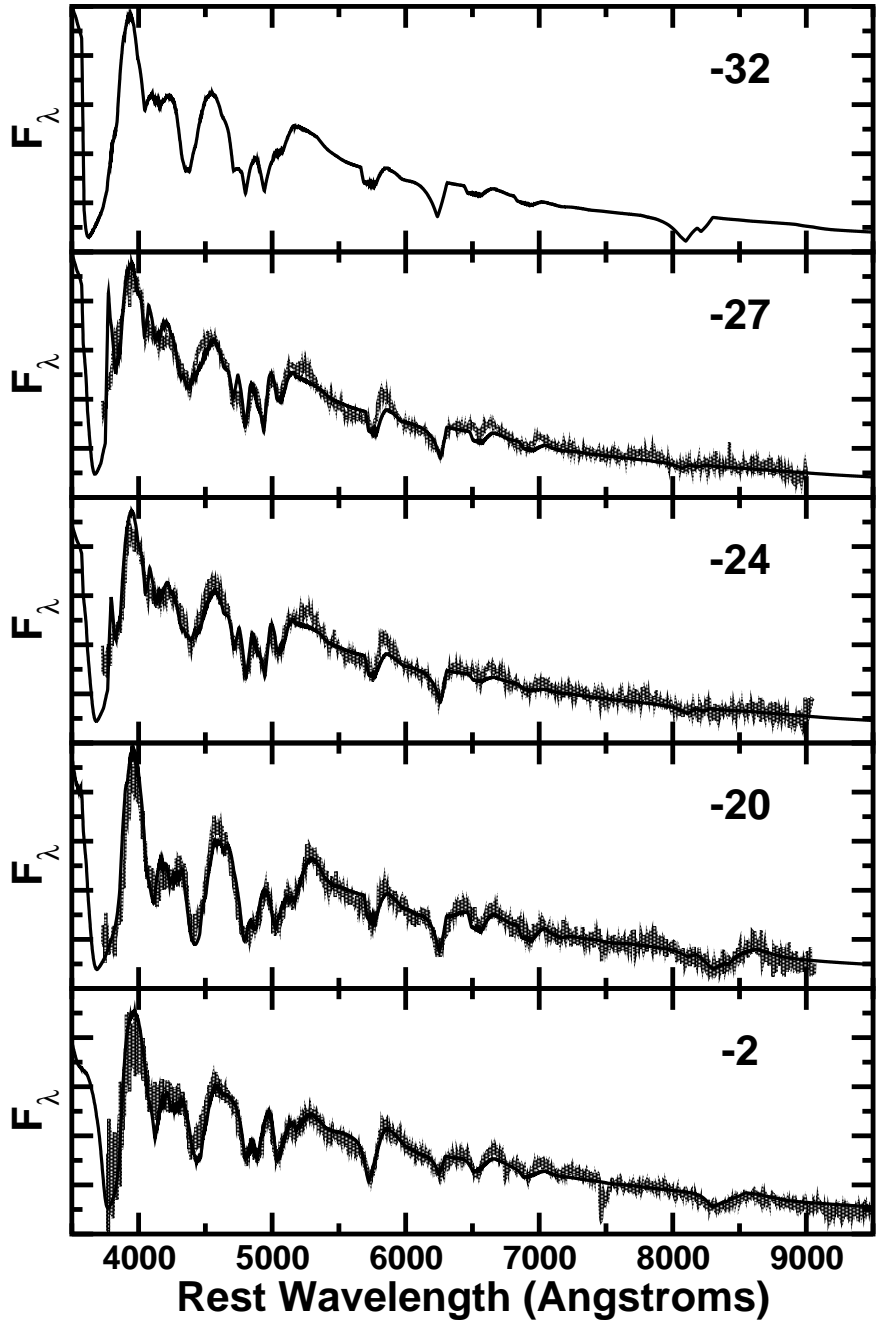}
\caption{SYNOW fits to the pre-maximum spectra of 
  SN~2005bf. The SYNOW synthetic spectra, which are described in the
  text, are overplotted on the observed
  spectra. The epoch is given in rest-frame days relative to
  $L_{\mathrm {bol}}$ maximum. \label{fig:synow_fits}} 
\end{figure}

\noindent As a tool for analyzing the pre-maximum spectral evolution
of SN~2005bf, we used the SYNOW code \citep{fisher00} to calculate
synthetic spectral fits to our first five spectra.  Following the 
precepts of \citet{branch02}, we assumed a power-law radial density
gradient of index $n = 8$ and a Boltzmann excitation temperature of
$T_{exc} = 7000$ K for all of these fits.  In general, only the
\ion{Fe}{2}, \ion{Ca}{2}, \ion{He}{1}, and \ion{H}{1} ions were 
included in the calculations. The fits were not intended to reproduce
all of the observed features but to provide support to the present
analysis. The addition of other ions would not modify our
conclusions. Figure~\ref{fig:synow_fits} shows
the final SYNOW fits overplotted on the observed spectra.

\begin{figure}
\epsscale{1.0}
\plotone{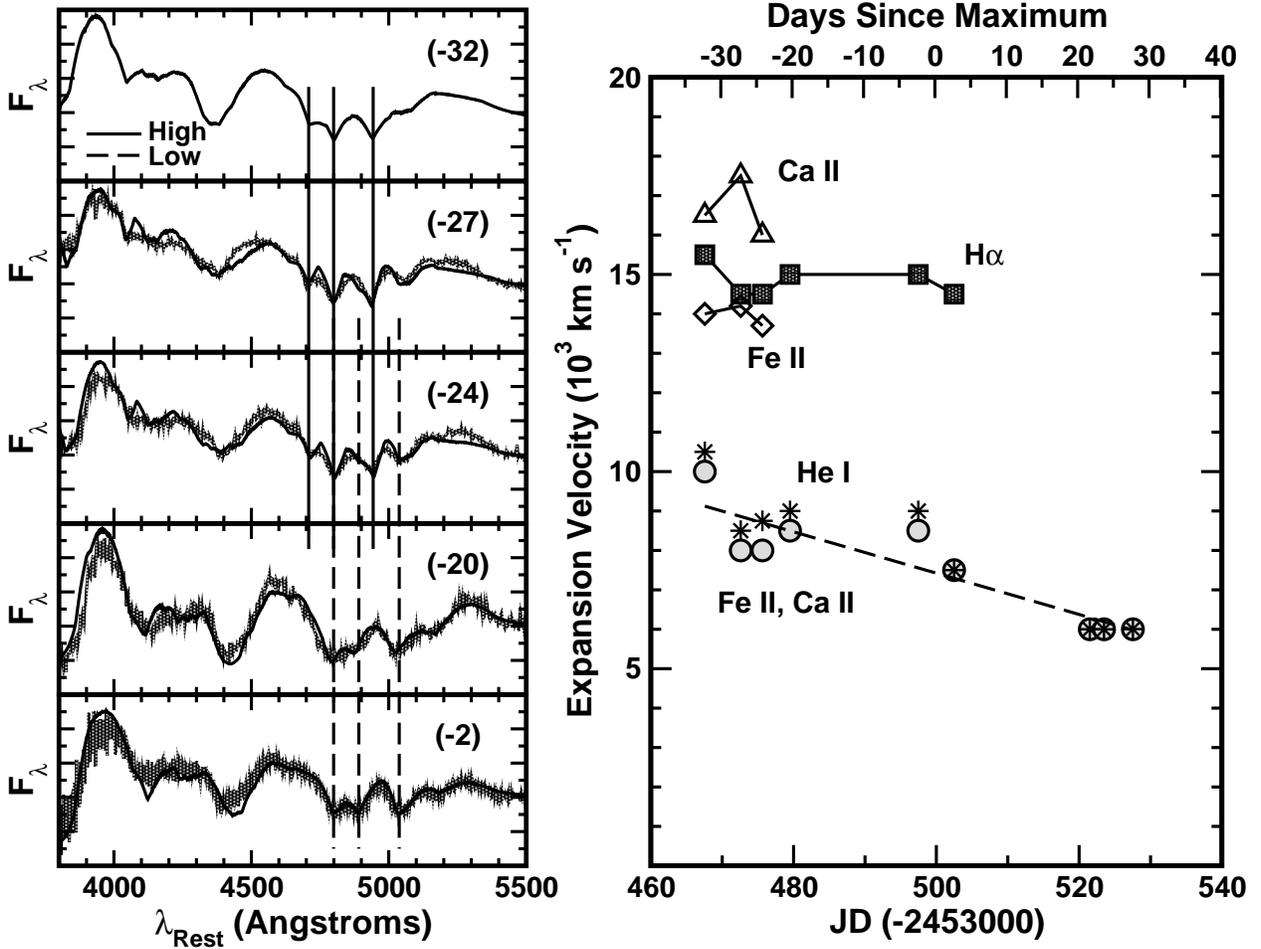}
\caption{(left) Evolution of the spectral region centered on the 
  \ion{Fe}{2} $\lambda$$\lambda$4924, 5018, 5169 lines during the rise 
  to maximum light.  The observed spectra are compared with 
  SYNOW synthetic spectral calculations.  The three vertical solid 
  lines indicate the approximate wavelengths 
  of this \ion{Fe}{2} multiplet in the high-velocity 
  gas which is visible during the peculiar initial maximum in 
  the $u'g'BV$ light curves.  The three vertical dashed lines show 
  the same multiplet at lower expansion velocity. Note that from day
  $-27$ through 
  day $-20$, both velocity components are clearly present in the spectra.
  (right) Expansion velocity measurements for SN~2005bf.  Values 
  were estimated from SYNOW analyses of the observed spectra, and have 
  typical errors of 500--1000 km s$^{-1}$.  The diamonds, triangles, and
  squares correspond to the high-velocity \ion{Fe}{2}, \ion{Ca}{2}, and
  H$\alpha$ absorption, respectively, 
  whereas the circles show the lower-velocity
  \ion{Fe}{2} and \ion{Ca}{2} which we associate with the photosphere.
  The expansion velocities of the \ion{He}{1} lines are plotted with
  an asterisk. The time axes are given in the observer's
  frame.\label{fig:vels}} 
\end{figure}

\noindent With the aid of the SYNOW fits we noticed that the
conversion from a Type~Ic to a Type~Ib SN was 
accompanied by a sudden change in the expansion velocity of the
absorption components of the \ion{Fe}{2}
$\lambda$$\lambda$4924, 5018, 5169 lines and the  
\ion{Ca}{2} $\lambda$$\lambda$8498, 8542, 8662 triplet feature.
The left half of Figure~\ref{fig:vels} shows that two separate 
components of the \ion{Fe}{2} lines were present in the 
spectrum from day $-24$: one at an expansion velocity of
$\sim$14,000 km s$^{-1}$, and the other at $\sim$8000 km s$^{-1}$.
In fact, low-velocity and high-velocity components of \ion{Fe}{2} and
\ion{Ca}{2} were identifiable as early as our first spectrum (day
$-32$) with the aid of the SYNOW fits.
At early epochs, a low-velocity \ion{Ca}{2}
component, weaker than its high velocity counterpart, helps to
fit the red wing of the H~\&~K absorption. Similarly, a
low-velocity \ion{Fe}{2} component added to the high-velocity one
improves the fit in the region between 4500 \AA\ and 5200 \AA.
The high-velocity component dominated the spectrum obtained on day
$-32$, but had disappeared by day $-2$. The spectra from day $-2$ onward
showed exclusively the low-velocity absorption component of \ion{Fe}{2}
and \ion{Ca}{2}. This strongly suggests that the 
appearance of high-velocity \ion{Fe}{2} and \ion{Ca}{2} lines is
associated with the initial maximum exhibited by the $u'g'BV$ light
curves. The evolution of the expansion velocities of these
components is summarized in the right half of Figure~\ref{fig:vels}.

\noindent During the period covered by our spectroscopic observations,
we find that the low-velocity component of \ion{Fe}{2} and \ion{Ca}{2} 
decreased roughly monotonically between $\sim$10,000 and $\sim$6000 km
s$^{-1}$. We have identified this with the photospheric velocity in
our SYNOW fits. Assuming this interpretation is correct, and that
the expansion was spherically symmetric, the 
\ion{He}{1} lines are most consistent with arising from a shell 
which has a maximum velocity of $\sim$11,000 km s$^{-1}$, 
whereas the high-velocity \ion{Fe}{2}, \ion{Ca}{2}, and H$\alpha$
lines are produced in a detached shell at velocities $\ge$14,000 km
s$^{-1}$. 

\noindent The 6250~\AA~absorption, like the high-velocity \ion{Fe}{2}
and \ion{Ca}{2} lines, was present at early epochs but disappeared by
the time of maximum light. A similar feature has been seen in several
Type~Ib SNe and its identification is a long-debated issue \citep[e.g.,
see][]{wheeler94,branch02}.  We consider the most likely
identification to be high-velocity ($\sim$15,000 km s$^{-1}$)
H$\alpha$ for two basic reasons: (1) the coincidence in expansion
velocity with the \ion{Fe}{2} and \ion{Ca}{2} lines which show a
similar association with the peculiar initial maximum in the $u'g'BV$
light curves, and (2) the presence of a very weak absorption feature
in the spectra from days $-24$ and $-20$ (marked with a dotted line in
Figure~\ref{fig:spec}) which could be identified as H$\beta$ at the
same expansion velocity and with a strength consistent with that
predicted by SYNOW spectra\footnote{According to \citet{wang05},
weak absorption consistent with H$\beta$ and H$\gamma$ was also
present in a spectrum obtained on Apr~30 (day $-9$).}.  We reject the
alternative identification of this feature with \ion{Si}{2}
$\lambda$6355 since it would imply an expansion velocity of $\sim$4800
km s$^{-1}$, which is significantly lower than the photospheric
velocity estimated at these early epochs\footnote{Nevertheless, we
cannot rule out that the absorption is produced by a blend of
high-velocity H$\alpha$ with \ion{Si}{2} $\lambda$6355 at 8000--10,000
km s$^{-1}$.}.  We consider identifications with \ion{C}{2}
$\lambda$6580 or \ion{Ne}{1} $\lambda$6402 unlikely for the same
reasons given by \citet{branch02} for ordinary Type~Ib SNe. Thus, a small
amount of hydrogen appears to be present in the outer ejecta of
SN~2005bf.

\subsection{Bolometric Light Curve}
\label{sec:bol}

\noindent The $u'g'r'i'BV$ photometry was used to compute a
quasi-bolometric light curve covering the wavelength range longward of
3000~\AA.  Our broad-band magnitudes were corrected for Galactic
extinction using $E(B-V)_{\mathrm {Gal}} = 0.045$ mag
\citep{schlegel98} and the reddening law with $R_V = 3.1$ given by
\citet{cardelli89}. Additional extinction originating in the host
galaxy is difficult to estimate.  Examination of our spectra from days
$-2$ and $+2$, which have the best wavelength resolution and
signal-to-noise ratios, reveals the presence of interstellar
absorption lines of 
\ion{Na}{1} D1 \& D2, and \ion{Ca}{2} H \& K at the redshift of the
host galaxy, suggesting that some extinction exists. However, the
uncertainties in estimating $E(B-V)_{\mathrm {Host}}$ from the
equivalent widths (EWs) of these lines are large. We measure the total
EW of the \ion{Na}{1} D1 \& D2 lines to be $\sim$1 \AA. According to
\citet{turatto03}, two values of the color excess would be favored:
$E(B-V)_{\mathrm {Host}}\approx 0.1$ and $0.5$ mag, depending on the
gas-to-dust ratio of the host galaxy environment. We consider the
value $E(B-V)_{\mathrm {Host}}\approx 0.5$ mag to be unrealistic
because it would imply color temperatures $\gtrsim$40,000 K during the
time of 
initial maximum (see below). Such high temperatures do not agree with
the ionic species observed in our spectra. Hence, the lower value of
$E(B-V)_{\mathrm {Host}}\approx 0.1$ mag seems more likely.  Close
examination of the day $-2$ spectrum reveals \ion{Na}{1}~D absorption
due to our own Galaxy at an equivalent width of approximately
$\sim$0.5~\AA.  If the interstellar medium in the host galaxy of
SN~2005bf is similar to that of the Galaxy, this would also suggest an
extinction of $E(B-V)_{\mathrm {Host}} \approx 0.1$ mag. Given the
uncertainties, we adopt $E(B-V)_{\mathrm {Host}}=0.0$ mag for our
analysis, but present the case of $E(B-V)_{\mathrm {Host}}=0.1$ mag as
an alternative.

\noindent The extinction corrected $u'g'r'i'BV$ broad-band magnitudes
were converted into monochromatic fluxes at the effective wavelengths
of 3557, 4825, 6261, 7672, 4448, and 5505 \AA, respectively, as given
by \citet[][their Tables 2a and 2c]{fukugita96}. On epochs where a
certain filter observation was not available, its magnitude was
interpolated in time from the light curve using the surrounding points
and a low-order polynomial. The total ``uvoir'' flux $F_{u'\rightarrow
i'}$ in the region between the effective wavelengths of the $u'$ and
$i'$ filters was integrated using the trapezoid approximation. On the
blue side of 3557 \AA\, the flux $F_{\nu}$(3557) derived from the $u'$
magnitude was extrapolated with a straight line to zero flux at 3000
\AA\ (the blueward limit of the $u'$ filter), and no significant flux
was supposed to be emitted at shorter wavelengths, an assumption based
on observations of other Type~Ib/Ic SNe and of SN~1987A
\citep{panagia03}. On the infrared side the flux was extrapolated to 
$\lambda = \infty$ using a BB model obtained by fitting
the $g'r'i'BV$ fluxes with a Planck function (shifted to the rest
frame of the SN)\footnote{The $u'$ points were excluded from the BB
fits because they clearly departed from the model, especially before
maximum light. The $u'$ fluxes generally lay below the fitted BB
curves, probably due to strong line blanketing in that part of the
spectrum. The effect of including $u'$ would be to lower the
derived BB temperature values. For consistency, we kept the same
policy for post-maximum epochs, even though there was a better
agreement of the $u'$ fluxes with the fitted BB curves.}.  The
integrated flux under the fitted Planck function between the effective
wavelength of the $i'$ filter and $\lambda = \infty$ was taken as the
IR correction. This correction remained below 35\% of the total flux
until a few days after maximum light and increased to 60\% thereafter.
A by-product of this procedure is the color temperature ($T_{bb}$)
obtained from the BB fits. The values of $T_{bb}$ for $E(B-V)_{\mathrm
{Host}}=0.0$ mag are given in Table~\ref{tab:phot} and shown in
Figure~\ref{fig:tcol} along with the observed SN colors\footnote{As a
further corroboration, we included the few NIR photometry points in
the BB fits. This exercise showed an agreement within 10\% between the
bolometric flux based on $u'$ through $i'$ and that based on $u'$
through $K_s$ (or $u'$ through $H$ on JD~2,453,483.5).  The BB fits
showed a systematic increase in temperature of about 500 to 800 K when
going from a range of $B$ through $i'$ to $B$ through $K_s$.}.  The
sum of $F_{u'\rightarrow i'}$ and the UV and IR corrections yielded
the bolometric flux $F_{\mathrm {bol}}$.

\begin{figure}
\epsscale{0.95}
\plotone{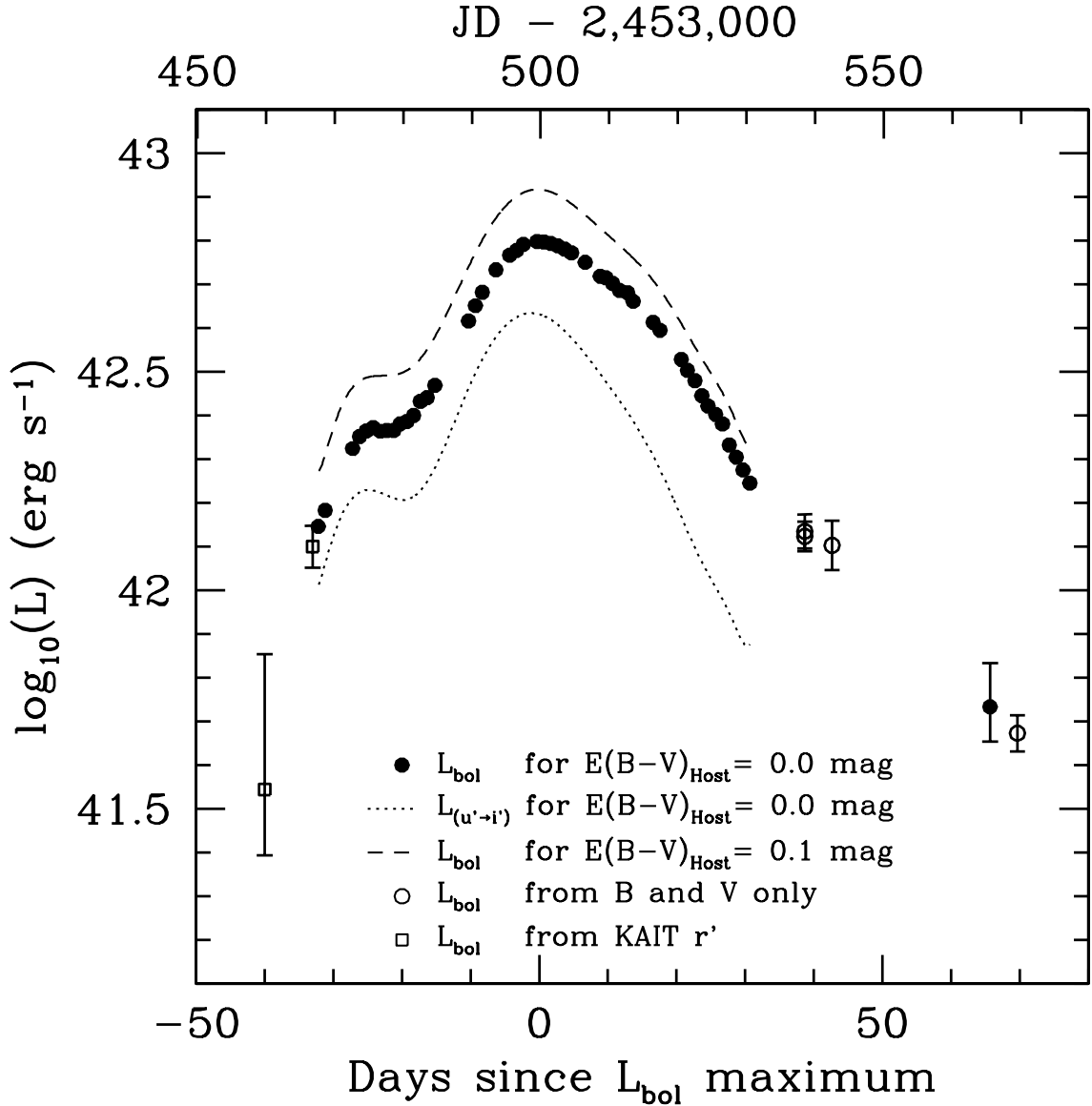}
\caption{Bolometric light curve of SN~2005bf as derived from the
  optical photometry. Filled circles mark bolometric luminosities
  computed from integrated $u'g'r'i'BV$ fluxes,
  plus UV and IR corrections, $E(B-V)_{\mathrm {Gal}}=0.045$ mag,
  $E(B-V)_{\mathrm {Host}}=0.0$, and an assumed distance of 83.8
  Mpc. Open circles mark bolometric luminosities derived from observed 
  $(B-V)$ colors and bolometric corrections. Open squares mark 
  luminosities derived from $r'$ magnitudes measured on the discovery
  images of KAIT, and an extrapolation of the bolometric
  correction. The dashed line shows a smooth-curve fit to the bolometric
  luminosities obtained assuming $E(B-V)_{\mathrm {Host}}=0.1$ mag.
  The dotted line shows a smooth-curve fit to the uvoir luminosities
  $L_{u'\rightarrow i'}$ derived in the range of $u'$ through
  $i'$, and assuming $E(B-V)_{\mathrm {Host}}=0.0$ mag.\label{fig:lbol}}  
\end{figure}

\noindent Both $F_{u'\rightarrow i'}$ and $F_{\mathrm {bol}}$ were
then transformed into uvoir ($L_{u'\rightarrow i'}$) and bolometric
luminosity ($L_{\mathrm {bol}}$), respectively. We assumed spherical
symmetry and used a
distance of 83.8$\pm$10.2 Mpc based on the Hubble law, a value of the
Hubble constant of $H_0 = 72 \pm 8$ km s$^{-1}$ Mpc$^{-1}$
\citep{freedman01}, and a recession velocity in the cosmic microwave
background frame of 6032 $\pm$ 300 km s$^{-1}$ (obtained from the
heliocentric velocity given by \citet{falco99}, and using the
velocity transformation tool from the NASA/IPAC Extragalactic
Database, NED.) The resulting
luminosities computed with $E(B-V)_{\mathrm {Host}}=0.0$ mag are
listed in Table~\ref{tab:phot}, and the bolometric light curves of
SN~2005bf are shown in Figure~\ref{fig:lbol} for both $E(B-V)_{\mathrm
{Host}}=0.0$ mag (filled circles) and $E(B-V)_{\mathrm {Host}}=0.1$
mag (dashed line). The uvoir luminosities are shown with a dotted
line. The luminosity corresponding to JD~2,453,565.5 is plotted with an
error bar between $4.6 \times 10^{41}$ and $5.4\times 10^{41}$ erg
s$^{-1}$ owing to the large uncertainties in the $r'i'$ magnitudes.

\noindent Four late-time observations obtained with the Swope 1-m
telescope between JD~2,453,538 and 2,453,570 involved only the $BV$
filters. In those cases, integration of the flux in the optical range
using the trapezoid approximation
was not possible. Instead, bolometric luminosities were
computed by determining bolometric corrections ($M_{\mathrm {bol}}-V$)
from $(B-V)$ colors. A linear relation with a scatter of $\sim$0.03
mag was found between these two quantities in the range $0.5
< (B-V) < 1.1$ mag. The $(B-V)$ colors for the late-time epochs lie in the
range from 1.04 to 1.18 mag. The relation above was used to derive the
$L_{\mathrm {bol}}$ values listed in parentheses in
Table~\ref{tab:phot}. Figure~\ref{fig:lbol} shows these values
with open circles and error bars derived by propagating the
uncertainties in $(B-V)$.

\noindent The observations obtained 30--70 days after
maximum bolometric luminosity indicate late-time decline rates of
$0.032 \pm 0.009$ mag day$^{-1}$ in $B$, $0.034 \pm 0.004$ mag day$^{-1}$
in $V$, and $0.038 \pm 0.002$ mag day$^{-1}$ in $M_{\mathrm {bol}}$. These
values are significantly larger than the decline rate of $0.0098$ mag
day$^{-1}$ predicted if all the energy from the decay of $^{56}$Co
into $^{56}$Fe was fully thermalized in the ejecta
\citep{woosley88}, implying that at this epoch the ejecta of
SN~2005bf did not trap all the $\gamma$-ray radiation. 
 
\noindent The discovery LOSS images from March 30 (JD~2,453,459.8)
and April 6 (JD~2,453,466.8) can be 
used to provide valuable information on the early rise of the bolometric
light curve. We consider here the case of $E(B-V)_{\mathrm 
{Host}}=0.0$ mag. The $r'$ 
magnitudes derived from the unfiltered images were converted to
bolometric magnitudes ($M_{\mathrm {bol}}$) by extrapolating an
average of the bolometric correction ($M_{\mathrm {bol}}-r'$) found
for the first ten days of follow-up observations with the Swope telescope
(JD~2,453,467.7 through 2,453,477.6) to the discovery epochs. The 
resulting bolometric luminosities are given in 
parentheses in Table~\ref{tab:phot}, and are plotted with open squares
in Figure~\ref{fig:lbol}. An uncertainty of $1.4\times10^{41}$ erg
s$^{-1}$ in the bolometric luminosity of
April 6 was estimated by summing in quadrature the photometric error
in $r'$ and an uncertainty of $0.1$ mag in the adopted bolometric
correction. In the case of March 30 the main source of error was
the extrapolation of the bolometric correction. The
uncertainty was estimated in this case by computing bolometric
luminosities for BB curves of varying temperatures in the range 3000
$<$ $T_{bb}$ $<$ 20,000 K, and passing through the $r'$ flux
point. This yielded an asymmetric error bar of $(2.50 < L_{\mathrm
  {bol}} < 7.15)\times 10^{41}$ erg s$^{-1}$.

\noindent A simple integration of the bolometric luminosity between 40
days before and 35 days after the time of maximum luminosity yields a
total bolometric energy output of $2.1\times 10^{49}$ erg.

\section{MODELS AND IMPLICATIONS}
\label{sec:disc}

\noindent Although SN~2005bf eventually evolved to resemble a fairly
typical Type~Ib SN,  
both its photometric and spectroscopic evolution leading up to maximum
light were unusual in many respects:

\begin{itemize}

\item The initial maximum observed in the $u'g'BV$ light curves
is, to our knowledge, without precedent in Type~Ib/Ic light curves.
Although it superficially resembles the shock outbreaks observed 
for the Type~Ib/Ic SN~1999ex \citep{stritzinger02} and the Type~IIb
SN~1993J \citep{richmond94}, in both of those cases the initial rise 
occurred very quickly ($<$ 2 days) whereas the first maximum 
in SN~2005bf occurred at least two weeks following
outburst.  This much longer rise time argues strongly against it being
due to shock outbreak.

\item The bolometric light curve did not reach maximum until at least
40 days after outburst.  Such a long rise time seems to be unique,
although few Type~Ib/Ic SNe have been caught early  
enough to be completely sure of this. We are aware of only 
two other Type~Ic events, the ``hypernovae'' SN~1997ef and SN~1998bw, which 
exhibited $V$ light curves of comparable width although with shorter rise
times \citep{galama98,iwamoto00}. 

\item SN~2005bf was also unusually luminous for a Type~Ib/Ic SN, 
radiating $2.1 \times 10^{49}$ erg in the wavelength range longward of
3000 \AA\ during 
the first $\sim$75 days following outburst (assuming spherical
symmetry). This is nearly identical  
to the energy that we derived integrating the bolometric light curve
of the ``hypernova'' SN~1998bw \citep{patat01}.

\item The transformation from a Type~Ic SN at early epochs 
to a Type~Ib SN by maximum light is also unprecedented. However, we 
must point out that few Type~Ib/Ic SNe have been observed
spectroscopically at such early epochs. It should be noted that
if SN~2005bf had not been discovered until maximum light, it would
have been classified as a fairly typical Type~Ib SN.  This serves to
emphasize the importance of observing SNe as soon as possible after
outburst.

\item The initial maximum in the $u'g'BV$ light curves was 
accompanied by the presence in the spectrum of high velocity 
absorption lines of \ion{Fe}{2}, \ion{Ca}{2}, and \ion{H}{1}. This
absorption had disappeared by the time of the principal
maximum in the bolometric light curve, which argues for it being
physically associated with the mechanism responsible for the initial
maximum.

\item The photospheric velocity of 9000--10,000 km s$^{-1}$ observed
at the early epochs ($\sim$30 days before maximum) was unusually low
compared with typical Type~Ib SNe \citep{branch02}.

\item None
of the spectra presented here shows the narrow or
intermediate-width emission lines caused by an interaction between
the ejecta and any preceding slow circumstellar wind, as observed
in Type~IIn events \citep{schlegel90,filippenko97}.

\end{itemize}

\noindent The unique properties of SN~2005bf summarized above indicate
an unusual event, 
one having certain features in common with ordinary Type~Ib/c 
SNe (see \S~\ref{sec:spanal}), Type~IIb events like SN~1993J
\citep{filippenko93}, and highly energetic SNe such as those seen
in conjunction with GRBs 990425 \citep{galama98} and 030329
\citep{hjorth03,stanek03,matheson03}. Both the high 
luminosity at the main peak and the long rise time argue
for a more massive helium core and larger mass of $^{56}$Ni than
in ordinary Type Ib/Ic SNe. Assuming a light curve dominated at peak by
radioactive decay and complete trapping, to produce a luminosity of $6
\times 10^{42}$ erg s$^{-1}$ 40 days after the explosion requires
about 0.6 M$_{\odot}$ of $^{56}$Co (made as $^{56}$Ni in the
explosion). Such a large mass of $^{56}$Ni is not produced in
spherically symmetric explosions unless the explosion energy is very
high, much greater than the canonical 10$^{51}$ erg, and the density
gradient near the iron core is shallow, as in massive stars ($\gtrsim$25
M$_\odot$) \citep{woosley95}.

\noindent On the other hand, the comparative faintness of the light
curve during 
its first few days and the fact that the effective temperature was
{\sl increasing} at earlier times, rather than cooling, argue for a
compact progenitor, neither a red nor blue supergiant. A red
supergiant of any sort would be too bright and too hot initially and
would cool monotonically with time. A blue supergiant with a radioactive peak
around 40 days would be too faint on day 10 and would also have a
declining temperature at early times.

\noindent The high-velocity hydrogen (with no low-velocity
counterpart) shows that 
the star had lost most, but not all of its hydrogen envelope. The
surface layers are shock accelerated to the highest speeds and a low
mass of hydrogen would not mix extensively with the rest of the
star. It seems that the progenitor was a Wolf-Rayet star of spectral
class WN \citep{maeder94} with less than 0.1 M$_{\odot}$ of hydrogen
in its outer layers. 

\begin{figure}
\plotone{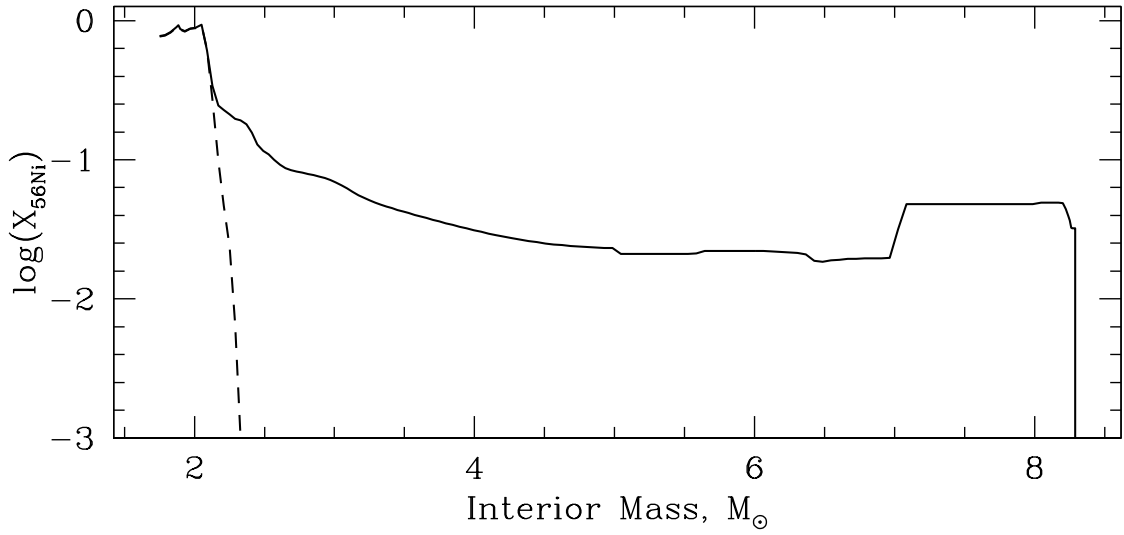}
\caption{The distribution of the $^{56}$Ni in the mixed ({\em solid
    line\,}) and unmixed ({\em dashed line\,}) versions of the $2
    \times 10^{51}$ erg explosion of an 8.29 M$_{\odot}$ WN
    star. The mass fraction of $^{56}$Ni is plotted as a function of
    the interior mass in the star.\label{fig:ni56}} 
\end{figure}

\noindent Within these confines, we explored a variety of models, but
found no 
physically reasonable solutions under the assumption of spherical
symmetry and a monotonically decreasing radial abundance of
$^{56}$Ni. The most successful one-dimensional
model was the $2.0 \times 10^{51}$
erg explosion of an 8.29 M$_{\odot}$ WN star with a mildly {\sl
inverted} distribution of $^{56}$Ni with mass as shown in
Figure~\ref{fig:ni56}. This star had 
a total $^{56}$Ni mass of 0.6 M$_{\odot}$, of which about 0.04
M$_{\odot}$ was artificially sited in the helium shell. Except for the
unusual ``mixing,'' this was essentially the same helium and 
heavy-element core one obtains by evolving a 25 M$_{\odot}$ main-sequence
star to its end point. The vast bulk of the hydrogenic envelope was
presumably lost to a binary companion quite late in the evolution.  We
did not conduct a broad survey of progenitor masses because we thought
the restriction of doing one-dimensional models did not warrant such an
approach. It is possible that an acceptable model might also have been
found for a higher-mass helium core and larger explosion
energy. 

\noindent Roughly the outer 0.05 M$_{\odot}$ of the WN star was
composed of hydrogen and helium with mass fractions of 0.34 and 0.66,
respectively. Terminal velocities ranged from 13,000 km s$^{-1}$ at
the base of this ``envelope'' to over 30,000 km s$^{-1}$ in the outer
0.001 M$_{\odot}$. The root-mean-square (rms)
velocity in the hydrogen-rich material was
18,000 km s$^{-1}$. All calculations of presupernova evolution and
explosive hydrodynamics up to the point of shock outbreak were
done with the KEPLER implicit hydrodynamics package
\citep{weaver78,woosley02}. The hydrogen envelope was removed at the end of
helium burning and the star was in hydrostatic and thermal equilibrium
at the time it exploded. 

\begin{figure}
\epsscale{1.1}
\plottwo{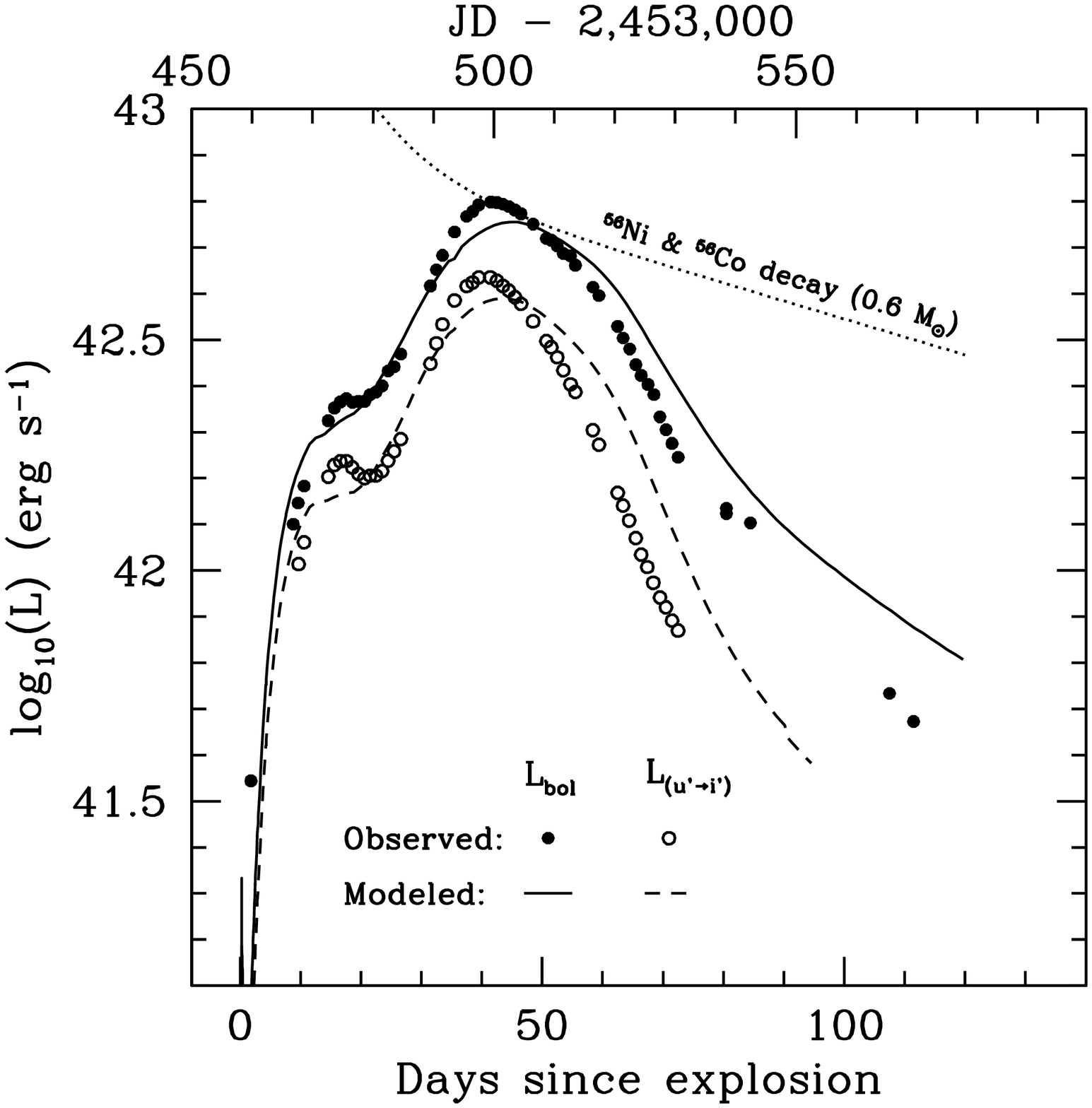}{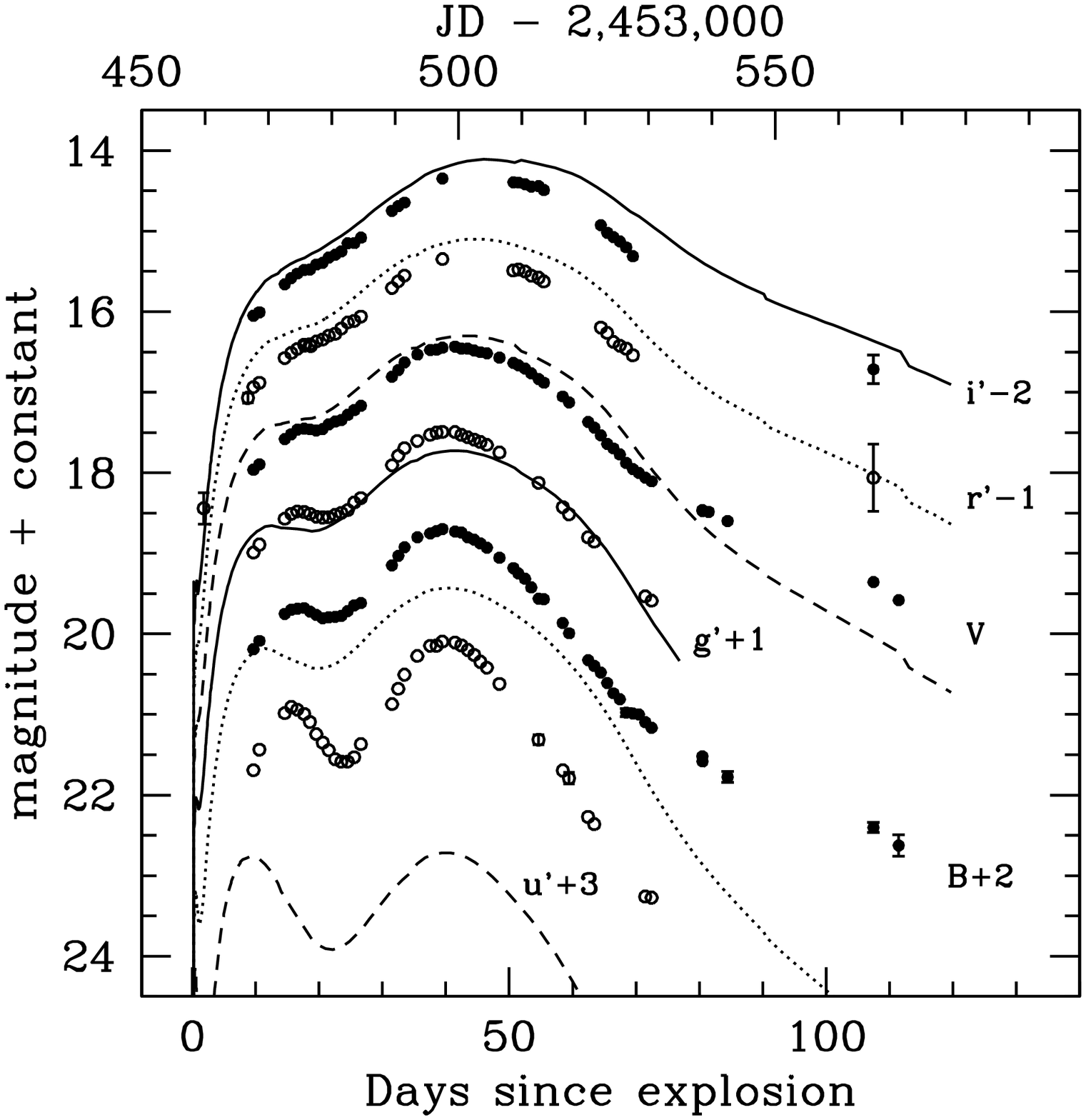}
\caption{(left) The modeled bolometric ({\em solid line\,}) and uvoir
  ({\em dashed line\,}) luminosities of the mixed 8.29 
  M$_{\odot}$ WN star explosion. The circles mark the uvoir luminosity
  $L_{u'\rightarrow i'}$ ({\em open\,}) and the bolometric luminosity
  $L_{\mathrm {bol}}$ ({\em closed\,}),
  as derived from the observations. The dotted line shows the
  predicted luminosity if the ejecta were able to trap the 
  entire gamma-ray output of 0.6 M$_\odot$ of
  radioactive $^{56}$Ni produced in the explosion. The time of
  outburst was assumed to be JD~2,453,458.
  (right) Modeled $u'g'r'i'BV$ light curves ({\em lines\,})
  compared with the observations ({\em circles\,}).\label{fig:modlc}}
\end{figure}

\noindent The light curve and multi-band photometry of this model were
then calculated using the STELLA code of \citet{blinnikov98} and
\citet{blinnikov00}. 
Though no fine tuning was attempted, Figure~\ref{fig:modlc} shows that
this model gave a qualitatively
good fit to both the observed bolometric luminosity and individual
colors up until the main peak. To 
achieve reasonable agreement beyond the main peak, however, much
greater gamma-ray leakage had to be invoked than was calculated by
STELLA for the spherically symmetric model.
This was achieved by 
turning down the gamma-ray opacity by a factor of 10 while leaving the
uvoir opacities unaltered. During the rise to maximum this alteration
had no effect since the optical depth was very large. After the
maximum though, the decline rate was greatly accelerated. We were
unable to find a model with unaltered gamma-ray opacity in which the
light curve peaked so late and yet declined so quickly.

\noindent All of these characteristics --- the high explosion energy,
large $^{56}$Ni mass, inverted distribution of $^{56}$Ni, and the need
for $\gamma$-ray leakage at late times --- are suggestive of a grossly
asymmetric explosion having many features in common with the
supernovae found coincident with GRBs
\citep[e.g.,][]{hoflich99,mazzali01,maeda03,mazzali05}. In fact,
SN~2005bf may be the best example so far of a ``transition object''
between the two classes of phenomena, namely GRBs and ordinary Type~Ib
SNe.

\noindent A realistic 
simulation of such an event will only be achievable in a multi-dimensional
simulation that captures the essence of energetic polar jets with
milder mass ejection in the equatorial plane.  Lacking the present
ability to do such a full multi-dimensional, relativistic simulation
including the necessary radiation transport, we also considered
multi-component models to reproduce the observed light curves. The
scenario we considered was that of a WN progenitor star of
about 6 to 15 M$_\odot$ whose iron core collapses either to a black
hole plus an accretion disk \citep{woosley93,macfayden99} or a very
rapidly rotating neutron star \citep{wheeler00}. A relativistic jet was
launched, as in current models for GRBs, but it was not observed,
possibly because it was beamed to other angles or because it had too
little energy in extremely relativistic ejecta. The jet was
accompanied, however, by vigorous $^{56}$Ni-rich 
outflows extending out to approximately $45^\circ$ \citep{macfayden99}.

\noindent The polar regions thus experienced a much more violent
explosion than the equator. A small fraction of the star's mass was
ejected with very 
high velocity at both poles along the star's rotational axis.  This
material contained about 0.1 M$_{\odot}$ of $^{56}$Ni per pole and the
observer was situated somewhat off axis.  The ensuing explosion can be
separated into two components: (1) a polar explosion containing a small
fraction of the total mass and moving at high velocity, and (2) the
explosion of the rest of the star. At first only the polar
explosion is observed, producing the initial maximum; when
that component fades and becomes transparent, the lower-velocity
ejecta become visible in the rise to the main maximum light.

\begin{figure}
\epsscale{1.1}
\plottwo{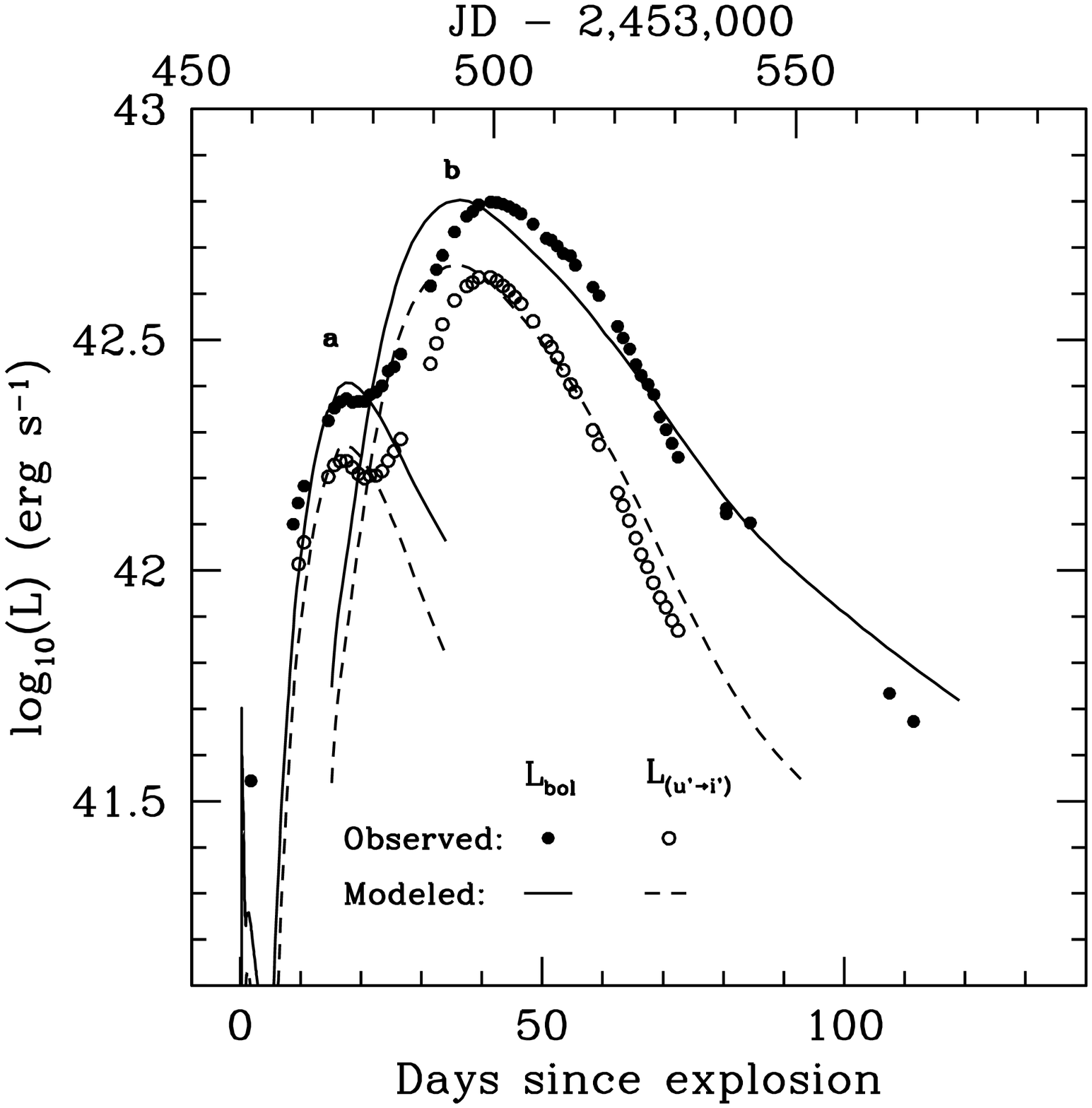}{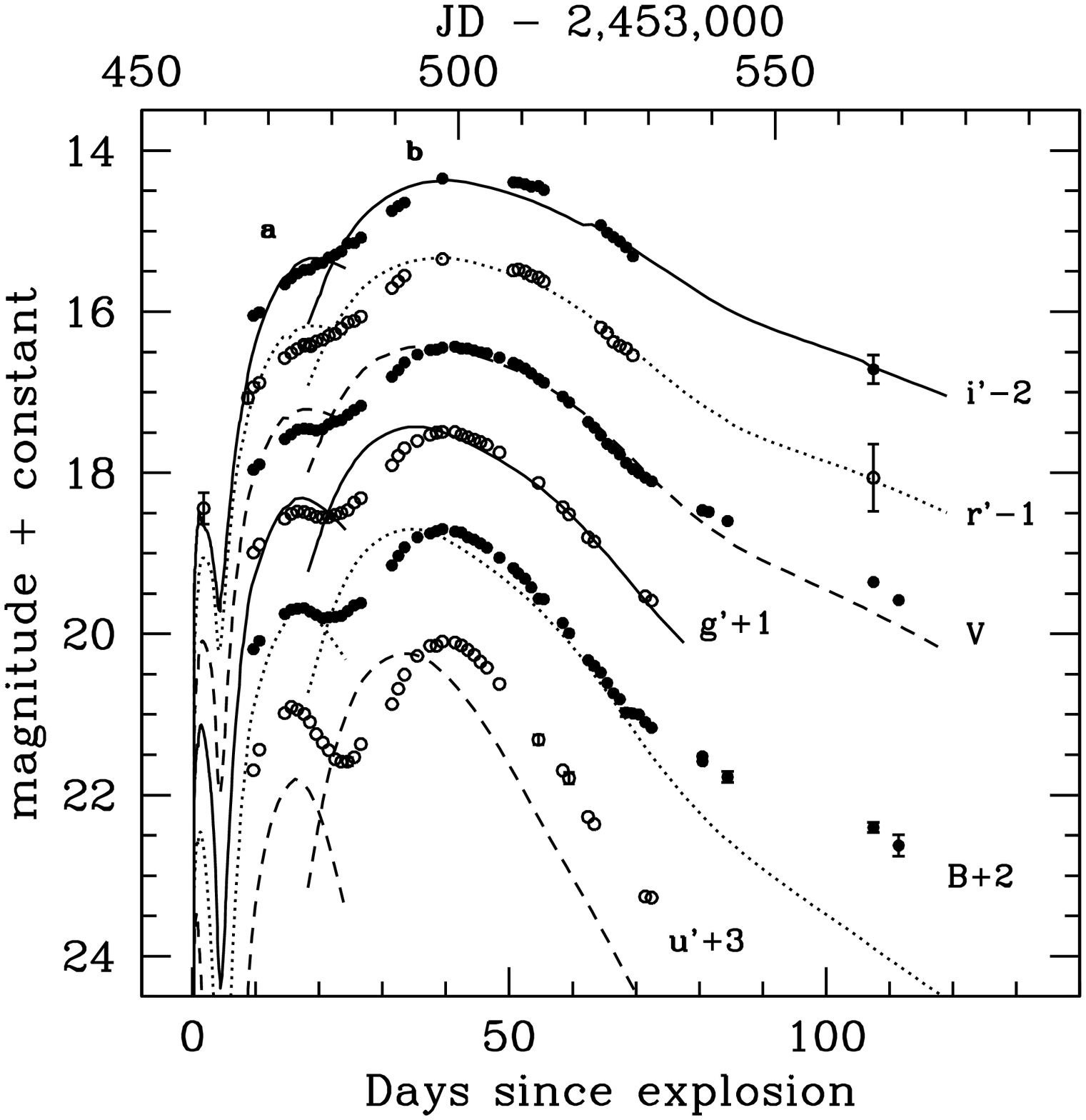}
\caption{(left) The bolometric ({\em solid line\,}) and uvoir
  ({\em dashed line\,}) luminosities of a composite model made of two
  explosions: (a) the $2 \times 10^{51}$ erg explosion of 1.72
  M$_{\odot}$ of helium and heavy elements containing 0.1 M$_{\odot}$
  of $^{56}$Ni, and (b) the $2 \times 10^{51}$ erg explosion of an 8.29
  M$_{\odot}$ WN star. Neither model had mixed $^{56}$Ni. The time of
  outburst was assumed to be JD~2,453,458. Component (a) is
  observable until about 20 days after outburst, when it is overrun by
  component (b). (right) The contributions of components (a) and (b) to
  the $u'g'r'i'BV$ light curves ({\em lines\,}) compared with the
  observations ({\em circles\,}).\label{fig:modlc2}} 
\end{figure}

\noindent As a specific example, by no means unique, we consider the
same 8.29 M$_{\odot}$ WN star as before, but with no mixing of the
$^{56}$Ni beyond where it was produced in the one-dimensional model
(the dashed 
line in Figure~\ref{fig:ni56}). In addition to this, we represent the
higher-velocity component by the $2 \times 10^{51}$ erg explosion of a
3.31 M$_{\odot}$ WR star that left behind a 1.52 M$_{\odot}$ remnant
(hence the $2 \times 10^{51}$ erg is concentrated in 1.79 M$_{\odot}$ of
ejecta). This component contains 0.1 M$_{\odot}$ of $^{56}$Ni
(presumably at each pole). The composite light curve and colors are
given in Figure~\ref{fig:modlc2}.  At early times ($t<20$ days),
before the rapidly expanding, low-mass 
component has become optically thin, the more slowly moving explosion is
occulted and blocked from view. At late times ($t>20$ days), the fast
component is invisible.  The geometry of the slower ejecta, viewed
along the equator (side),
resembles the number ``8.'' Gamma rays can
diffuse out much more effectively along the polar axis than along the
equator. To account for this, the gamma-ray opacity has been reduced
by a factor of 10.

\section{CONCLUSIONS}

\noindent The computer simulations presented here, though clearly
suffering from being one-di\-men\-sion\-al, 
strongly suggest that the light curve and the two-component 
spectrum of SN~2005bf can be qualitatively explained by the very
energetic explosion of a massive star that lost almost all of
its hydrogen envelope and exploded very asymmetrically. The energy and
$^{56}$Ni mass implied are large compared with those of common Type~Ib
SNe, but the energy is intermediate between ordinary supernovae
and ``hypernovae,'' suggesting that SN~2005bf might be some sort of
transition object between these two classes of phenomena. Given the
high energy, and the possible detection of polarization \citep{wang05},
it is very likely that the explosion was asymmetric and will
ultimately need to be modeled in at least two dimensions. The
$^{56}$Ni distribution could have been very aspherical, and this
particularly complicates the modeling of the first peak.

\noindent Given the previously observed association of GRBs and
energetic 
supernovae, it is quite plausible that SN~2005bf was powered by a
central engine like those being discussed for GRBs.  To retain the
necessary degree of rotation, it is more likely that the envelope of
the star was lost to a binary companion rather than in an ordinary
standard wind. The longer the star remains a red supergiant, the more
its core is braked by torques from the slowly rotating envelope
\citep{heger05,woosley06}. Two-dimensional simulations should be
conducted to verify the interpretation of this event as a possible 
``missing link.'' Radio observations to limit the amount of 
relativistic ejecta are also encouraged.

\acknowledgments 

We would like to thank David Branch for allowing us to
use SYNOW, and Jerod Parrent for providing us with the code. We thank
the referee for his/her valuable comments.
This material is based upon work supported by the National Science
Foundation (NSF) under grant AST--0306969. 
We also acknowledge support from {\it Hubble Space Telescope}
grant GO-09860.07-A from
the Space Telescope Science Institute, which is operated by
the Association of Universities for Research in Astronomy, Inc., under
NASA contract NAS 5-26555. 
SB is supported in part by the RFBR grants
05-02-17480, 04-02-16793. BLL was supported by an
NSERC PGS B and a Walter C. Sumner Fellowship.
MH acknowledges support
provided by NASA through Hubble Fellowship grant HST-HF-01139.01-A.
MH and GG are grateful for support from the
Centro de Astrof\'{\i}sica FONDAP 15010003. A.V.F.'s group at
U.C. Berkeley is supported by of NSF grant AST-0307894; he is
also grateful for a Miller Research Professorship, during which part of this
work was completed.  KAIT was made possible by generous donations from Sun
Microsystems, Inc., the Hewlett-Packard Company, AutoScope Corporation, Lick
Observatory, the NSF, the University of California,
and the Sylvia \& Jim Katzman Foundation. The W. M. Keck Observatory is
operated as a scientific partnership among the California Institute of
Technology, the University of California, and NASA; the Observatory was
made possible by the generous financial support of the W. M. Keck
Foundation. This research has made use of the NASA/IPAC Extragalactic
Database (NED) which is operated by the Jet Propulsion Laboratory,
California Institute of Technology, under contract with the National
Aeronautics and Space Administration.

\clearpage
\input{tab1.tex}

\clearpage
\input{tab2.tex}

\clearpage
\input{tab3.tex}

\clearpage
\input{tab4.tex}

\clearpage
\input{tab5.tex}

\clearpage
\input{tab6.tex}

\clearpage

\clearpage

\clearpage

\clearpage

\clearpage

\clearpage

\clearpage

\clearpage

\clearpage

\clearpage

\clearpage

%\clearpage
%\begin{figure}
%\plotone{}
%\caption{\label{}}
%\end{figure}

\end{document}

%% file: tab1.tex
\begin{deluxetable} {lcccccc}
\tabletypesize{\scriptsize}
\tablecolumns{7}
\tablewidth{0pt}
\tablecaption{Photometry of the comparison stars in the field of SN 2005bf\label{tab:stds}}
\tablehead{
\colhead{Star} & 
\colhead{} & 
\colhead{}   & 
\colhead{}   & 
\colhead{}   & 
\colhead{}   &
\colhead{}   \\
\colhead{ID} &
\colhead{$u'$} &
\colhead{$g'$} & 
\colhead{$r'$} & 
\colhead{$i'$} & 
\colhead{$B$} & 
\colhead{$V$}}
\startdata
C1 &  $\cdots$ &  $\cdots$ &  $\cdots$ &  $\cdots$ & 14.657(016) &  $\cdots$ \\ 
C2 & 16.360(021) & 15.186(015) & 14.762(015) & 14.630(015) & 15.545(009) & 14.943(010) \\ 
C3 & 19.203(016) & 16.529(008) & 15.463(009) & 15.067(009) & 17.141(009) & 15.963(008) \\ 
C4 & 17.105(016) &  $\cdots$ &  $\cdots$ &  $\cdots$ & 15.081(016) &  $\cdots$ \\ 
C5 & 19.988(034) & 17.292(009) & 16.115(008) & 15.624(008) & 17.938(010) & 16.675(007) \\ 
C6 & 16.254(020) & 15.147(011) & 14.753(011) & 14.603(011) & 15.476(009) & 14.905(010) \\ 
C7 & 21.254(205) & 18.924(009) & 18.013(008) & 17.676(008) & 19.446(016) & 18.434(008) \\ 
C8 & 18.328(013) & 17.040(008) & 16.529(008) & 16.328(008) & 17.433(009) & 16.743(007) \\ 
C9 & 22.883(797) & 20.755(028) & 19.362(018) & 18.302(009) & 19.504(477) & 19.957(024) \\ 
C10 & 19.416(033) & 18.473(009) & 18.125(008) & 18.022(013) & 18.770(011) & 18.259(008) \\ 
C11 & 20.418(046) & 18.628(008) & 17.953(008) & 17.691(008) & 19.104(015) & 18.247(008) \\ 
C12 & 19.604(044) & 17.845(008) & 17.176(008) & 16.939(008) & 18.326(015) & 17.472(007) \\ 
C13 & 21.969(371) & 20.692(053) & 19.294(017) & 17.730(010) & 21.872(192) & 19.940(029) \\ 
C14 & 17.251(019) & 15.996(008) & 15.520(008) & 15.353(008) & 16.371(009) & 15.709(007) \\ 
\enddata
\tablecomments{Uncertainties given in parentheses in thousandths of a
  magnitude correspond to the rms of the magnitudes obtained on four
  photometric nights, with a minimum uncertainty of 0.015 mag for an
  individual measurement.}
%\tablenotetext{a}{}
\end{deluxetable}

%% file: tab2.tex
%\documentclass[10pt,preprint]{aastex}
%\begin{document}
%Dummy line
%\pagestyle{empty}
%\ptlandscape

\begin{deluxetable} {lccccccccccc}
\tabletypesize{\scriptsize}
\rotate
\tablecolumns{12}
%\tablenum{1}
\tablewidth{0pt}
\tablecaption{$u'g'r'i'BV$ photometry of SN 2005bf\label{tab:phot}}
\tablehead{
\colhead{JD} &
\colhead{Epoch\tablenotemark{a}} &
\colhead{  } &
\colhead{  } &
\colhead{  } &
\colhead{  } &
\colhead{  } &
\colhead{  } &
\colhead{  } &
\colhead{$T_{bb}$} & 
\colhead{$\log L_{(u'\rightarrow i')}$} & 
\colhead{$\log L_{\mathrm {bol}}$} \\
\colhead{$-2,453,000$} &
\colhead{(days)} &
\colhead{Telescope} &
\colhead{$u'$} & 
\colhead{$g'$} & 
\colhead{$r'$} & 
\colhead{$i'$} & 
\colhead{$B$} & 
\colhead{$V$} & 
\colhead{(K)} &
\colhead{(erg s$^{-1}$)} & 
\colhead{(erg s$^{-1}$)}} 
%% Hay que corregir:
%% Agregar 5 noches KAIT. 
%% Poner Clay r' e i' del dia 497.5 separados del Swope. 
%% '0.000(000)' por '$\cdots$'
%% '-inf' por '$\cdots$'
%% '  0 &' por '$\cdots$ &'
%% '$-0$' por '$0$'
%% y ponerle parentesis a las L_bol de JD 459.79, 466.80, 538, 542 y 569.
\startdata
444.81 & $-54$ & KAIT &   $\cdots$ &   $\cdots$ &  $>20.0$ &   $\cdots$ &   $\cdots$ &   $\cdots$ &    $\cdots$ &    $\cdots$ &  $\cdots$ \\
459.79 & $-39$ & KAIT &   $\cdots$ &   $\cdots$ &  19.441(197) &   $\cdots$ &   $\cdots$ &   $\cdots$ &    $\cdots$ &    $\cdots$ &  (41.544) \\
466.80 & $-32$ & KAIT &   $\cdots$ &   $\cdots$ &  18.070(062) &   $\cdots$ &   $\cdots$ &   $\cdots$ &    $\cdots$ &    $\cdots$ &  (42.100) \\
467.65 & $-32$ & Swope &  18.691(021) &  17.991(015) &  17.935(015) &  18.048(015) &  18.191(016) &  17.962(015) &   9567 &  42.014 &  42.146 \\
468.58 & $-31$ & Swope &  18.436(016) &  17.891(015) &  17.881(015) &  18.006(017) &  18.086(016) &  17.894(015) &  10082 &  42.061 &  42.183 \\
472.59 & $-27$ & Swope &  17.983(016) &  17.571(015) &  17.575(015) &  17.659(015) &  17.750(016) &  17.581(015) &  10040 &  42.203 &  42.325 \\
473.61 & $-26$ & Swope &  17.908(016) &  17.508(015) &  17.513(015) &  17.582(015) &  17.701(016) &  17.520(015) &   9872 &  42.228 &  42.352 \\
474.57 & $-25$ & Swope &  17.939(016) &  17.478(015) &  17.462(015) &  17.526(015) &  17.686(016) &  17.464(015) &   9575 &  42.236 &  42.365 \\
475.64 & $-24$ & Swope &  17.994(016) &  17.484(015) &  17.409(015) &  17.481(015) &  17.678(016) &  17.451(015) &   9214 &  42.237 &  42.373 \\
475.78 & $-24$ & KAIT  & $\cdots$  & $\cdots$  &  17.419(062)  & $\cdots$   & $\cdots$  & $\cdots$  & $\cdots$  & $\cdots$  & $\cdots$  \\
476.56 & $-23$ & Swope &  18.092(016) &  17.510(015) &  17.406(015) &  17.477(015) &  17.721(016) &  17.460(015) &   8912 &  42.223 &  42.365 \\
476.73 & $-23$ & KAIT  & $\cdots$  & $\cdots$  &  17.426(033)  & $\cdots$   & $\cdots$  & $\cdots$  & $\cdots$  & $\cdots$  & $\cdots$  \\
477.56 & $-22$ & Swope &  18.244(016) &  17.543(015) &  17.371(015) &  17.414(015) &  17.762(016) &  17.475(015) &   8297 &  42.209 &  42.366 \\
478.57 & $-21$ & Swope &  18.348(017) &  17.554(015) &  17.344(015) &  17.392(015) &  17.806(016) &  17.457(015) &   7974 &  42.200 &  42.366 \\
479.52 & $-20$ & Swope &  18.443(023) &  17.550(015) &  17.298(015) &  17.328(015) &  17.792(016) &  17.397(015) &   7713 &  42.206 &  42.381 \\
480.54 & $-19$ & Swope &  18.554(020) &  17.524(015) &  17.275(015) &  17.294(015) &  17.787(016) &  17.365(015) &   7622 &  42.206 &  42.387 \\
481.51 & $-18$ & Swope &  18.587(026) &  17.496(015) &  17.208(015) &  17.253(015) &  17.777(016) &  17.346(015) &   7462 &  42.215 &  42.400 \\
482.52 & $-17$ & Swope &  18.583(050) &  17.462(030) &  17.134(026) &  17.150(020) &  17.715(025) &  17.282(022) &   7262 &  42.238 &  42.433 \\
483.54 & $-16$ & Swope &  18.530(037) &  17.370(015) &  17.113(015) &  17.148(015) &  17.646(016) &  17.223(015) &   7596 &  42.258 &  42.442 \\
484.61 & $-15$ & Swope &  18.369(041) &  17.314(015) &  17.060(015) &  17.082(015) &  17.616(016) &  17.166(015) &   7485 &  42.285 &  42.469 \\
489.51 & $-10$ & Swope &  17.871(016) &  16.905(015) &  16.704(015) &  16.746(015) &  17.148(016) &  16.808(015) &   8033 &  42.448 &  42.617 \\
490.50 & $-9$ & Swope &  17.681(016) &  16.787(015) &  16.622(015) &  16.688(015) &  17.029(016) &  16.722(015) &   8350 &  42.493 &  42.652 \\
491.48 & $-8$ & Swope &  17.506(016) &  16.694(015) &  16.552(015) &  16.645(017) &  16.923(016) &  16.630(015) &   8675 &  42.534 &  42.683 \\
493.51 & $-6$ & Swope &  17.273(043) &  16.603(015) &   $\cdots$ &   $\cdots$ &  16.799(025) &  16.532(015) &   8677 &  42.586 &  42.734 \\
495.53 & $-4$ & Swope &  17.147(016) &  16.528(015) &   $\cdots$ &   $\cdots$ &  16.754(016) &  16.477(015) &   8486 &  42.616 &  42.768 \\
496.54 & $-3$ & Swope &  17.145(017) &  16.501(015) &   $\cdots$ &   $\cdots$ &  16.724(016) &  16.472(015) &   8449 &  42.624 &  42.778 \\
497.46 & $-2$ & Swope &  17.097(016) &  16.493(015) &   $\cdots$ &   $\cdots$ &  16.697(016) &  16.448(015) &   8333 &  42.635 &  42.793 \\
497.47 & $-2$ & Clay  &  $\cdots$  &  $\cdots$  &  16.348(015) &  16.347(030)\tablenotemark{b} &  $\cdots$  &  $\cdots$  &   $\cdots$     &  $\cdots$ &  $\cdots$ \\ 
499.45 & $ 0$ & Swope &  17.105(016) &  16.495(015) &   $\cdots$ &   $\cdots$ &  16.730(016) &  16.432(015) &   8051 &  42.636 &  42.799 \\
500.46 & $ 1$ & Swope &  17.142(016) &  16.529(015) &   $\cdots$ &   $\cdots$ &  16.743(016) &  16.460(015) &   7873 &  42.628 &  42.798 \\
501.46 & $ 2$ & Swope &  17.198(016) &  16.559(015) &   $\cdots$ &   $\cdots$ &  16.801(016) &  16.460(015) &   7601 &  42.617 &  42.794 \\
502.46 & $ 3$ & Swope &  17.261(016) &  16.586(015) &   $\cdots$ &   $\cdots$ &  16.829(016) &  16.483(015) &   7466 &  42.607 &  42.789 \\
503.46 & $ 4$ & Swope &  17.346(016) &  16.619(015) &   $\cdots$ &   $\cdots$ &  16.875(016) &  16.502(015) &   7298 &  42.592 &  42.781 \\
504.48 & $ 5$ & Swope &  17.418(016) &  16.653(015) &   $\cdots$ &   $\cdots$ &  16.928(016) &  16.518(015) &   7141 &  42.578 &  42.773 \\
506.47 & $ 7$ & Swope &  17.620(016) &  16.749(015) &   $\cdots$ &   $\cdots$ &  17.055(016) &  16.571(015) &   6798 &  42.540 &  42.751 \\
508.65 & $ 9$ & Swope &   $\cdots$ &   $\cdots$ &  16.489(015) &  16.393(015) &  17.182(016) &  16.636(015) &   6592 &  42.497 &  42.720 \\
509.46 & $ 9$ & Swope &   $\cdots$ &   $\cdots$ &  16.474(015) &  16.397(015) &  17.250(016) &  16.664(015) &   6371 &  42.484 &  42.716 \\
510.50 & $11$ & Swope &   $\cdots$ &   $\cdots$ &  16.501(015) &  16.418(015) &  17.317(023) &  16.706(017) &   6225 &  42.462 &  42.703 \\
511.48 & $11$ & Swope &   $\cdots$ &   $\cdots$ &  16.550(015) &  16.447(015) &  17.423(016) &  16.766(015) &   6048 &  42.434 &  42.687 \\
512.65 & $13$ & Swope &  18.313(062) &  17.123(015) &  16.579(015) &  16.441(015) &  17.561(018) &  16.835(015) &   5719 &  42.404 &  42.682 \\
513.46 & $13$ & Swope &   $\cdots$ &   $\cdots$ &  16.626(015) &  16.490(015) &  17.570(016) &  16.881(015) &   5776 &  42.387 &  42.662 \\
516.45 & $16$ & Swope &  18.695(024) &  17.426(015) &   $\cdots$ &   $\cdots$ &  17.863(016) &  17.052(015) &   5383 &  42.304 &  42.614 \\
517.45 & $17$ & Swope &  18.792(071) &  17.513(029) &   $\cdots$ &   $\cdots$ &  17.992(016) &  17.127(015) &   5240 &  42.273 &  42.596 \\
520.46 & $20$ & Swope &  19.271(035) &  17.801(015) &   $\cdots$ &   $\cdots$ &  18.324(016) &  17.367(015) &   4945 &  42.169 &  42.529 \\
521.44 & $21$ & Swope &  19.361(043) &  17.854(022) &   $\cdots$ &   $\cdots$ &  18.399(016) &  17.439(015) &   4938 &  42.141 &  42.504 \\
522.45 & $22$ & Swope &   $\cdots$ &   $\cdots$ &  17.195(015) &  16.924(015) &  18.478(025) &  17.535(018) &   4879 &  42.109 &  42.480 \\
523.47 & $23$ & Swope &   $\cdots$ &   $\cdots$ &  17.263(015) &  17.018(015) &  18.609(018) &  17.641(015) &   4821 &  42.070 &  42.446 \\
524.45 & $24$ & Swope &   $\cdots$ &   $\cdots$ &  17.377(020) &  17.075(027) &  18.737(017) &  17.697(015) &   4750 &  42.034 &  42.423 \\
525.45 & $25$ & Swope &   $\cdots$ &   $\cdots$ &  17.421(015) &  17.127(015) &  18.811(038) &  17.773(027) &   4698 &  42.007 &  42.403 \\
526.45 & $26$ & Swope &   $\cdots$ &   $\cdots$ &  17.458(015) &  17.201(015) &  18.975(051) &  17.878(018) &   4575 &  41.974 &  42.381 \\
527.51 & $27$ & Swope &   $\cdots$ &   $\cdots$ &  17.540(015) &  17.313(015) &  18.987(016) &  17.953(015) &   4688 &  41.942 &  42.333 \\
528.47 & $28$ & Swope &   $\cdots$ &   $\cdots$ &   $\cdots$ &   $\cdots$ &  18.999(016) &  18.003(015) &   4737 &  41.920 &  42.305 \\
529.48 & $29$ & Swope &  20.260(046) &  18.535(015) &   $\cdots$ &   $\cdots$ &  19.097(017) &  18.061(015) &   4713 &  41.891 &  42.275 \\
530.47 & $30$ & Swope &  20.274(039) &  18.588(021) &   $\cdots$ &   $\cdots$ &  19.167(018) &  18.105(017) &   4729 &  41.870 &  42.245 \\
538.48 & $38$ & Swope &   $\cdots$ &   $\cdots$ &   $\cdots$ &   $\cdots$ &  19.520(037) &  18.464(015) &    $\cdots$ &    $\cdots$ &  (42.124) \\
538.49 & $38$ & Swope &   $\cdots$ &   $\cdots$ &   $\cdots$ &   $\cdots$ &  19.583(043) &  18.472(015) &    $\cdots$ &    $\cdots$ &  (42.135) \\
539.46 & $39$ & Swope &   $\cdots$ &   $\cdots$ &   $\cdots$ &   $\cdots$ &   $\cdots$ &  18.484(019) &    $\cdots$ &    $\cdots$ &    $\cdots$ \\
542.46 & $42$ & Swope &   $\cdots$ &   $\cdots$ &   $\cdots$ &   $\cdots$ &  19.776(068) &  18.598(018) &    $\cdots$ &    $\cdots$ &  (42.103) \\
565.47 & $64$ & Swope &   $\cdots$ &   $\cdots$ &  19.061(416) & 18.715(177) &  20.403(063) &  19.357(022) &   4685 &    41.279  &  41.734 \\
569.47 & $68$ & Swope &   $\cdots$ &   $\cdots$ &   $\cdots$ &   $\cdots$ &  20.625(133) &  19.582(042) &    $\cdots$ &    $\cdots$ &  (41.673) \\
\enddata
\tablecomments{Photometric uncertainties given in parentheses in
  thousandths of a magnitude. A minimum uncertainty of 0.015 mag was
  set.}
\tablenotetext{a}{Rest-frame days since $L_{\mathrm {bol}}$ maximum
  (JD~2,453,499.8).}
\tablenotetext{b}{Synthetic magnitude computed from the spectrum obtained
  the same night.}
\end{deluxetable}
%\end{document}

%% file: tab3.tex
\begin{deluxetable} {lcc}
%
% The values here were extracted from the *.log files in
% ./analysis/datemax/
%
\tabletypesize{\scriptsize}
\tablecolumns{3}
\tablewidth{0pt}
\tablecaption{Light-curve parameters for SN 2005bf\label{tab:lpar}}
\tablehead{
\colhead{} & 
\colhead{JD at Peak} &  
\colhead{Peak}\\
\colhead{Filter} & 
\colhead{$-2,453,000$} &  
\colhead{Magnitude}}
\startdata
$u'$ & 497.9 $\pm$ 0.5 & \phs17.10 $\pm$ 0.01 \\
$g'$ & 498.1 $\pm$ 0.5 & \phs16.50 $\pm$ 0.01 \\
$r'$ & 500.2 $\pm$ 1.0 & \phs16.32 $\pm$ 0.01 \\
$i'$ & 502.9 $\pm$ 1.0 & \phs16.32 $\pm$ 0.02 \\
$B$  & 497.4 $\pm$ 0.5 & \phs16.71 $\pm$ 0.02 \\
$V$  & 498.8 $\pm$ 0.5 & \phs16.44 $\pm$ 0.01 \\[4pt]
$M_{\mathrm {bol}}$ & 499.8 $\pm$ 0.5 & $-$18.32 $\pm$ 0.03\tablenotemark{a} \\
\enddata
\tablecomments{Uncertainties in the peak magnitudes were estimated
  from the rms of the photometric points about a low-order polynomial
  fit. Uncertainties in the dates of maximum are based on the
  data sampling rates.} 
\tablenotetext{a}{Absolute bolometric magnitude based on $L_{\mathrm
    {bol}}$ from Table~\ref{tab:phot} (no host-galaxy extinction
  correction applied).} 
\end{deluxetable}

%% file: tab4.tex
\begin{deluxetable} {lcccc}
\tabletypesize{\scriptsize}
\tablecolumns{5}
\tablewidth{0pt}
\tablecaption{Near-infrared photometry of two comparison stars\label{tab:nirst}}
\tablehead{
\colhead{Star} & 
\colhead{} & 
\colhead{}   & 
\colhead{}   & 
\colhead{}   \\
\colhead{ID} &
\colhead{$Y$} &
\colhead{$J$} & 
\colhead{$H$} & 
\colhead{$K_s$}}
\startdata
C10 &  17.40(02) &  17.16(02) &  16.87(05) &  16.72(06) \\ 
C11 &  16.97(02) &  16.58(02) &  16.12(02) &  16.00(03) \\ 
\enddata
\tablecomments{Uncertainties given in parentheses in hundredths of a
  magnitude. A minimum uncertainty of 0.02 mag was set.}
%\tablenotetext{a}{}
\end{deluxetable}

%% file: tab5.tex
\begin{deluxetable} {lcccccc}
\tabletypesize{\scriptsize}
\tablecolumns{7}
\tablewidth{0pt}
\tablecaption{Near-infrared photometry of SN 2005bf\label{tab:nir}}
\tablehead{
\colhead{JD} & 
\colhead{Epoch\tablenotemark{a}} & 
\colhead{}   & 
\colhead{}   & 
\colhead{}   & 
\colhead{}   & \\
\colhead{$-2,453,000$} &
\colhead{(days)} &
\colhead{$Y$} & 
\colhead{$J$} & 
\colhead{$H$} & 
\colhead{$K_s$} & 
\colhead{Instrument}}
\startdata
483.51 & $-16$    & 16.93(03) & 16.75(03) & 16.65(05) &  $\cdots$ & WIRC \\ 
483.52 & $-16$    & 16.92(03) & 16.76(03) & 16.64(05) &  $\cdots$ & WIRC \\ 
511.46 & \phs$11$ & 16.19(03) & 15.95(03) & 15.87(05) & 15.80(06) & PANIC \\ 
538.46 & \phs$38$ & $\cdots$  & 16.91(03) &  $\cdots$ & 16.43(05) & PANIC \\ 
\enddata
\tablecomments{WIRC data were obtained with two detectors and their
  magnitudes agree within uncertainties. We use the averages in the
  present analysis. Uncertainties are given in parentheses in
  hundredths of a magnitude. A minimum uncertainty of 0.02 mag was set
  for individual measurements. Uncertainties in the zero points
  arising from uncertainties in the magnitudes of comparison stars
  were added in quadrature to the measurement errors.}
\tablenotetext{a}{Rest-frame days since $L_{\mathrm {bol}}$ maximum
  (JD~2,453,499.8).}
\end{deluxetable}

%% file: tab6.tex
\begin{deluxetable} {lcccccc}
\tabletypesize{\scriptsize}
\tablecolumns{7}
\tablewidth{0pt}
\tablecaption{Spectroscopic observations of SN 2005bf\label{tab:spec}}
\tablehead{
\colhead{JD} & 
\colhead{Epoch\tablenotemark{a}} & 
\colhead{ } & 
\colhead{Wavelength} & 
\colhead{Resolution\tablenotemark{b}} & 
\colhead{Exposure} & 
\colhead{$\langle S/N \rangle$\tablenotemark{c}}\\
\colhead{$-2,453,000$} & 
\colhead{(days)} & 
\colhead{Instrument} & 
\colhead{Range (\AA)} &
\colhead{(\AA)} & 
\colhead{(s)} & 
\colhead{(in 10 \AA)}}
\startdata
467.57 & $-32$     & WFCCD  & 3800 -- \phn9235  & 8   & 900  & 69  \\
472.60 & $-27$     & WFCCD  & 3800 -- \phn9235  & 8   & 900  & 88  \\
475.63 & $-24$     & WFCCD  & 3800 -- \phn9235  & 8   & 900  & 74  \\
479.51 & $-20$     & WFCCD  & 3800 -- \phn9235  & 8   & 700  & 32  \\
497.47 & \phn$-2$  & LDSS-3 & 3838 -- 10000     & 4\tablenotemark{d} & 200  & 262 \\
501.89 & \phn\phs2 & LRIS   & 3100 -- 9350      & 4   & 300/360\tablenotemark{e}  & 343 \\
502.48 & \phn\phs3 & LDSS-3 & 6057 -- 10000     & 5   & 900  & 200 \\
521.50 & \phs21    & WFCCD  & 3800 -- \phn8125  & 8\tablenotemark{f}  & 1200 & 64  \\
523.49 & \phs23    & WFCCD  & 3800 -- \phn8128  & 8   & 1200 & 42  \\
527.54 & \phs27    & WFCCD  & 3800 -- \phn8128  & 8   & 600  & 55  \\
\enddata
\tablecomments{Some spectra are the combination of multiple
  observations. In those cases, total exposure times are given.}
\tablenotetext{a}{Rest-frame days since $L_{\mathrm {bol}}$ maximum
  (JD~2,453,499.8).} 
\tablenotetext{b}{Average resolution obtained from the FWHM of arc-lamp lines.}
\tablenotetext{c}{Average signal-to-noise ratio in 10 \AA\ bins
  calculated in the range from 4000 to 8000 \AA.}
\tablenotetext{d}{Resolution of 2.9 \AA\ in the blue channel ($3800 <
  \lambda < 6000$ \AA), and 4.7 \AA\ in the red channel ($6000
  <\lambda < 10,000$ \AA).}
\tablenotetext{e}{Different exposure times in blue channel (300 s) and
  red channel (360 s).}
\tablenotetext{f}{A wide slit ($8.''65$) was used. In this case the
  resolution was estimated from the FWHM of the spatial profile of the SN.}
\end{deluxetable}

%% file: ms.bbl
\begin{thebibliography}{}

\bibitem[Allington-Smith(1994)]{allington94}
  Allington-Smith,~J. et~al. 1995, \pasp, 106, 983

\bibitem[Bessell(1990)]{bessell90} Bessell,~M.~S. 1990, \pasp, 102, 1181

\bibitem[Bessell(1999)]{bessell99} Bessell,~M.~S. 1999, \pasp, 111, 1426

\bibitem[Blinnikov et al.(1998)]{blinnikov98} Blinnikov,~S.~I.,
  Eastman,~R., Bartunov,~O.~S., Popolitov,~V.~A., \& Woosley,~S.~E.
  1998, \apj, 496, 454 

\bibitem[Blinnikov \& Sorokina(2000)]{blinnikov00} Blinnikov,~S.~I., \&
  Sorokina,~E.~I. 2000, \aap, 356, L30. 

\bibitem[Branch et al.(2002)]{branch02} Branch,~D., et~al. 2002, \apj,
  566, 1005 

\bibitem[Cardelli, Clayton, \& Mathis(1989)]{cardelli89}
  Cardelli,~J.~A., Clayton,~G.~C., \& Mathis,~J~.S. 1989, \apj, 345, 245

\bibitem[Clocchiatti \& Wheeler(1997)]{clocchiatti97}
  Clocchiatti,~A., \& Wheeler, J.~C. 1997, \apj, 491, 375

\bibitem[Fisher (2000)]{fisher00} Fisher,~A. 2000, Ph.D. thesis, 
  Univ. Oklahoma

\bibitem[Falco et al.(1999)]{falco99} Falco,~E.~E., et~al. 1999, PASP,
  111, 438 

\bibitem[Filippenko(1982)]{filippenko82} Filippenko,~A.~V. 1982, PASP, 94,
715

\bibitem[Filippenko(1997)]{filippenko97} Filippenko,~A.~V. 1997,
   ARAA, 35, 309

\bibitem[Filippenko(2005)]{filippenko05} Filippenko,~A.~V. 2005, in The Fate
of the Most Massive Stars, ed. R. Humphreys \& K. Stanek (San Francisco:
ASP), 33

\bibitem[Filippenko et al.(1993)]{filippenko93} Filippenko,~A.~V.,
  Matheson,~T., \& Ho,~L.~C. 1993, \apjl, 415, L103 
 
\bibitem[Filippenko et al.(1995)]{filippenko95} Filippenko,~A.~V.,
  et~al. 1995, \apjl, 450, L11  

\bibitem[Filippenko et al.(2001)]{filippenko01} Filippenko,~A.~V.,
  Li,~W., Treffers,~R.~R., \& Modjaz,~M. 2001, in Small-Telescope
  Astronomy on Global Scales, ed. W.-P. Chen, C. Lemme, \&
  B. Paczy\'{n}ski (San Francisco: ASP), 121 

\bibitem[Foley et al.(2003)]{foley03} Foley,~R.~J., et~al. 2003, PASP,
  115, 1220 
 
\bibitem[Freedman et al.(2001)]{freedman01} Freedman,~W.~L.,
  et~al. 2001, \apj, 553, 47 

\bibitem[Fukugita et al.(1996)]{fukugita96} Fukugita,~M.,
  Ichikawa,~T., Gunn,~J.~E., Doi,~M., Shimasaku,~K., \&
  Schneider,~D.~P. 1996, \aj, 111, 1748

\bibitem[Galama et al.(1998)]{galama98} Galama, T.~J., et~al. 1998,
  Nature, 395, 670 

%\bibitem[Hamuy et al.(1988)]{hamuy88} Hamuy,~M., Suntzeff,~N.~B.,
%  Gonz\'alez,~R., \& Martin,~G. 1994, \aj, 95, 63

\bibitem[Hamuy(2004)]{hamuy04} Hamuy,~M. 2004, in Stellar Collapse,
  ed. C.~L. Fryer (Dordrecht: Kluwer), 39

\bibitem[Hamuy et al.(2005)]{hamuy05} Hamuy,~M., Contreras,~C.,
  Gonzalez,~S., \& Krzeminski,~W. 2005, IAU Circ., 8520

\bibitem[Hamuy et al.(1994)]{hamuy94} Hamuy,~M., Suntzeff,~N.~B.,
  Heathcote,~S.~R., Walker,~A.~R., Gigoux,~P., Phillips,~M.~M. 1994,
  \pasp, 106, 566

\bibitem[Hamuy et al.(2002)]{hamuy02} Hamuy,~M., et~al. 2002, \aj,
  124, 417

\bibitem[Hamuy et al.(2006)]{hamuy06} Hamuy,~M., et~al. 2006, \pasp,
  in press (Paper I)

\bibitem[Harkness et al.(1987)]{harkness87} Harkness,~R.~P.,
  et~al. 1987, \apj, 317, 355 

\bibitem[Heger, Woosley, \& Spruit(2005)]{heger05} Heger,~A.,
  Woosley,~S.~E., \& Spruit,~H.~C. 2005, \apj, 626, 350 

\bibitem[Hjorth et al.(2003)]{hjorth03} Hjorth,~J., et~al. 2003, \nat,
  423, 847 

\bibitem[H\"oflich, Wheeler, \& Wang(1999)]{hoflich99} H\"oflich,~P.,
  Wheeler,~J.~C., \& Wang,~L. 1999, \apj, 521, 179 

\bibitem[Iwamoto et al.(2000)]{iwamoto00} Iwamoto,~K., et~al. 2000, 
  \apj, 534, 660

\bibitem[Landolt(1992)]{landolt92} Landolt,~A.~U. 1992, \aj, 104, 340

\bibitem[Li et al.(2000)]{li00} Li,~W., et~al. 2000,
in Cosmic Explosions, ed. S. S. Holt, \& W. W. Zhang (New York: AIP), 103

\bibitem[MacFadyen \& Woosley(1999)]{macfayden99} MacFadyen,~A.~I., \&
  Woosley,~S.~E.\ 1999, \apj, 524, 262 

\bibitem[Maeda et al.(2003)]{maeda03} Maeda,~K., Mazzali,~P.~A.,
  Deng,~J., Nomoto,~K., Yoshii,~Y., Tomita,~H., \& Kobayashi,~Y.\
  2003, \apj, 593, 931 

\bibitem[Maeder \& Meynet(1994)]{maeder94} Maeder,~A., \& Meynet,~G.
  1994, \aap, 287, 803 

\bibitem[Martini et al.(2004)]{martini04} Martini,~P., Persson,~S.~E.,
  Murphy,~D.~C., Birk,~C., Shectman,~S.~A., Gunnels,~S.~M., \&
  Koch,~E. 2004, SPIE, 5492, 1653

\bibitem[Matheson et al.(2003)]{matheson03} Matheson,~T., et~al.
   2003, ApJ, 599, 394

\bibitem[Mazzali et al.(2001)]{mazzali01} Mazzali,~P.~A., Nomoto,~K.,
  Patat,~F., \& Maeda,~K.\ 2001, \apj, 559, 1047

\bibitem[Mazzali et al.(2005)]{mazzali05} Mazzali,~P.~A., et~al.
  2005, Science, 308, 1284 
   
\bibitem[Modjaz, Kirshner, \& Challis(2005)]{modjaz05} Modjaz,~M.,
  Kirshner,~R., \& Challis,~P. 2005, IAU Circ., 8509 

\bibitem[Monard, Moore, \& Li(2005)]{monard05} Monard,~L.~A.~G.,
  Moore,~M., \& Li,~W. 2005, IAU Circ., 8507 
 
\bibitem[Morrell et al.(2005)]{morrell05} Morrell,~N., Hamuy,~M.,
  Folatelli,~G., \& Contreras,~C. 2005, IAU Circ., 8509 

\bibitem[Oke et al.(1995)]{oke95} Oke,~J.~B., et~al. 1995, PASP,
   107, 375

\bibitem[Panagia(2003)]{panagia03} Panagia,~N. 2003, in Lecture Notes
  in Physics, 598, Supernovae and
  Gamma Ray Bursters, ed. K.~W. Weiler (Berlin: Springer), 113

\bibitem[Patat et al.(2001)]{patat01} Patat,~F., et~al. 2001, \apj,
  555, 900
 
\bibitem[Persson et al.(1998)]{persson98} Persson,~S.~E. et~al. 1998,
  \aj, 116, 2475 

\bibitem[Persson et al.(2002)]{persson02} Persson,~S.~E.,
  Murphy,~D.~C., Gunnels,~S.~M., Birk,~C., Bagish,~A., \&
  Koch,~E. 2002, \aj, 124, 619 

%\bibitem[Pun et al.(1995)]{pun95} Pun, C.~S.~J., et al. 1995, \apjs,
%  99, 223

%\bibitem[Rauscher et al.(2002)]{rauscher02} Rauscher, T., Heger, A.,
%  Hoffman, R.~D., \& Woosley, S.~E. 2002, \apj, 576, 323

\bibitem[Richmond et al.(1994)]{richmond94} Richmond,~M.~W., et~al.
 1994, \aj, 107, 1022

\bibitem[Schlegel(1990)]{schlegel90} Schlegel,~E.~M. 1990, \mnras,
  244, 269 

\bibitem[Schlegel, Finkbeiner, \& Davis(1998)]{schlegel98}
 Schlegel,~D.~J., Finkbeiner,~D.~P., \& Davis,~M. 1998, \apj, 500,
 525 

\bibitem[Smith et al.(2002)]{smith02} Smith,~J.~A., et~al. 2002, \aj,
  123, 2121

\bibitem[Stanek et al.(2003)]{stanek03} Stanek,~K.~Z., et~al. 2003,
  \apj, 599, L95

\bibitem[Stanek et al.(2005)]{stanek05} Stanek,~K.~Z., et~al. 2005,
  \apj, 626, L5

\bibitem[Stritzinger et al.(2002)]{stritzinger02} Stritzinger,~M., 
  et~al. 2002, \aj, 124, 2100

\bibitem[Turatto, Benetti, \& Cappellaro(2003)]{turatto03}
  Turatto,~M., Benetti,~S., \& Cappellaro,~E. 2003, in 
  From Twilight to Highlight:
  the Physics of Supernovae, ed. B. Leibundgut \& W. Hillebrandt
  (Berlin: Springer-Verlag), 200

\bibitem[Wang \& Baade(2005)]{wang05} Wang,~L., \& Baade,~D. 2005,
  IAU Circ., 8521

\bibitem[Weaver, Zimmerman, \& Woosley(1978)]{weaver78} Weaver,~T.~A.,
  Zimmerman,~G.~B., \& Woosley,~S.~E. 1978, \apj, 225, 1021

\bibitem[Wheeler et al.(1994)]{wheeler94} Wheeler,~J.~C.,
  Harkness,~R.P., Clocchiatti,~A., Benetti,~S., Brotherton,~M.~S.,
  DePoy,~E.~L., \& Elias,~J. 1994, \apj, 436, L135

\bibitem[Wheeler et al.(2000)]{wheeler00} Wheeler,~J.~C., Yi,~I.,
  H\"oflich,~P., \& Wang,~L. 2000, \apj, 537, 810 

\bibitem[Woosley(1988)]{woosley88}  Woosley,~S.~E. 1988, \apj, 330,
  218 

\bibitem[Woosley(1993)]{woosley93} Woosley,~S.~E. 1993, \apj, 405,
  273 
 
\bibitem[Woosley \& Heger(2006)]{woosley06} Woosley,~S.~E., \&
  Heger,~A. 2006, submitted to \apj, astro-ph/0508175  

\bibitem[Woosley et al.(2002)]{woosley02} Woosley,~S.~E., Heger,~A.,
  \& Weaver,~T.~A. 2002, Reviews of Modern Physics, 74, 1015

\bibitem[Woosley~\& Weaver(1995)]{woosley95} Woosley,~S.~E., \&
  Weaver, T.~A. 1995, \apjs, 101, 181 

%\bibitem[Woosley, Heger, \& Hoffman(2005)]{woosley05b} Woosley, S. E.,
%  Heger, A., \& Hoffman, R. D. 2005, in preparation for \apj

%\bibitem[Zhang, Woosley, \& Heger(2004)]{zhang04} Zhang, W., Woosley,
%  S.~E., \& Heger, A. 2004, \apj, 608, 365 

%\bibitem[]{1052} Author, I. year, Journal, Vol, page

\end{thebibliography}
